\documentclass[twocolumn]{aastex631}

\usepackage{amsmath}        %for specific mathematical equations
\usepackage{gensymb}        %for degrees
\usepackage{fixltx2e}

\begin{document}

\title{Redshifted iron emission and absorption lines in the Chandra X-ray spectrum of Centaurus A}

\author[0000-0002-5924-4822]{David Bogensberger}
\affiliation{Department of Astronomy, The University of Michigan, 1085 South University Avenue, Ann Arbor, MI 48103, USA}

\author[0000-0003-2869-7682]{Jon Miller}
\affiliation{Department of Astronomy, The University of Michigan, 1085 South University Avenue, Ann Arbor, MI 48103, USA}

\author[0000-0002-0273-218X]{Elias Kammoun}
\affiliation{IRAP, Universit\'{e} de Toulouse, CNRS, UPS, CNES, 9, Avenue du Colonel Roche, BP 44346, F-31028, Toulouse Cedex 4, France}

\affiliation{INAF -- Osservatorio Astrofisico di Arcetri, Largo Enrico Fermi 5, I-50125 Firenze, Italy
}

\affiliation{Dipartimento di Matematica e Fisica, Universit\`{a} Roma Tre, via della Vasca Navale 84, I-00146 Rome, Italy}

\author{Richard Mushotzky}
\affiliation{Department of Astronomy, University of Maryland, College Park, MD 20742, USA}

\author{Laura Brenneman}
\affiliation{Center for Astrophysics, Observatory Building E, 60 Garden St, Cambridge, MA 02138, USA}

\author{W. N. Brandt}
\affiliation{Department of Astronomy and Astrophysics, 525 Davey Lab, The Pennsylvania State University, University Park, PA 16802, USA}
\affiliation{Institute for Gravitation and the Cosmos, The Pennsylvania State University, University Park, PA 16802, USA}
\affiliation{Department of Physics, 104 Davey Laboratory, The Pennsylvania State University, University Park, PA 16802, USA}

\author[0000-0002-8294-9281]{Edward M. Cackett}
\affiliation{Department of Physics and Astronomy, Wayne State University, 666 W.\ Hancock St, Detroit, MI, 48201, USA}

\author{Andrew Fabian}
\affiliation{Institute of Astronomy, University of Cambridge, Madingley Road, Cambridge, CB3 0HA, United Kingdom}

\author{Jelle Kaastra}
\affiliation{SRON Netherlands Institute for Space Research, Niels Bohrweg 4, 2333 CA Leiden, Netherlands}
\affiliation{Leiden Observatory, Leiden University, PO Box 9513, 2300 RA Leiden, The Netherlands}

\author{Shashank Dattathri}
\affiliation{Department of Astronomy, Yale University, Kline Tower
266 Whitney Avenue, New Haven, CT 06511, USA}

\author{Ehud Behar}
\affiliation{Faculty of Physics, Technion - Israel Institute of Technology, Haifa, 3200003, Israel}

\author{Abderahmen Zoghbi}
\affiliation{Department of Astronomy, University of Maryland, College Park, MD 20742}
\affiliation{HEASARC, Code 6601, NASA/GSFC, Greenbelt, MD 20771}
\affiliation{CRESST II, NASA Goddard Space Flight Center, Greenbelt, MD 20771}

%\email{\href{mailto:dbogen@umich.edu}{dbogen@umich.edu}}



\begin{abstract}
\noindent
Cen A hosts the closest active galactic nucleus to the Milky Way, which makes it an ideal target for investigating the dynamical processes in the vicinity of accreting supermassive black holes. In this paper, we present 14 \emph{Chandra} HETGS spectra of the nucleus of Cen A that were observed throughout 2022. We compared them with each other, and contrasted them against the two previous \emph{Chandra} HETGS spectra from 2001. This enabled an investigation into the spectral changes occurring on timescales of months and 21 years. All \emph{Chandra} spectra could be well fitted by an absorbed power law with a strong and narrow Fe K$\alpha$ line, a leaked power law feature at low energies, and Si and S K$\alpha$ lines that could not be associated with the central engine. The flux of the continuum varied by a factor of $2.74\pm0.05$ over the course of the observations, whereas the Fe line only varied by $18.8\pm8.8\%$. The photon index increased over 21 years, and the Hydrogen column density varied significantly within a few months as well. The Fe K$\alpha$ line was found at a lower energy than expected from the Cen A redshift, amounting to an excess velocity of $326^{+84}_{-94}~\mathrm{km}~\mathrm{s}^{-1}$ relative to Cen A. We investigated warped accretion disks, bulk motion, and outflows as possible explanations of this shift. The spectra also featured ionized absorption lines from Fe XXV and Fe XXVI, describing a variable inflow. 
\end{abstract}

\keywords{Active Galactic Nuclei (16) --- X-ray astronomy(1810) --- Spectroscopy(1558) --- Black holes(162)}


\section{Introduction} \label{sec:intro}

\noindent
The distance to the galaxy Centaurus A (commonly abbreviated as Cen A, and also known as NGC 5128) is merely $3.8 \pm 0.1 ~ \mathrm{Mpc}$ \citep{2010PASA...27..457H}. It hosts the nearest active galactic nucleus (AGN) of type Seyfert 2 \citep{2011A&A...531A..70B}, which is powered by a supermassive black hole (SMBH) with a mass of $5.5\pm3.0\times10^7~\mathrm{M}_{\odot}$ \citep{2009MNRAS.394..660C, 2022ApJS..261....2K}. \citet{1970ApJ...161L...1B} detected X-ray emission from Cen A for the first time, and it has subsequently played a crucial role in the developing scientific understanding of AGNs. 

AGN X-ray spectra are commonly described by an absorbed power law, with a strong Fe K$\alpha$ emission line, and reflection features such as the Compton hump. Other features include a soft excess, and additional, but weaker emission or absorption lines, which can trace the composition, ionization, and dynamics of the system.

To better understand the processes at work in the central engine of an AGN, we decided to investigate how X-ray spectral features vary over the course of months and decades in the AGN of Cen A. AGNs feature large flux variability on these timescales, which is mostly associated with changes in the accretion rate, the absorbing column density \citep{2002ApJ...571..234R}, or the power law slope \citep{2016MNRAS.459.3963C}. Some AGNs have strongly variable column densities, which change between Compton-thin and Compton-thick regimes, and are known as changing-look AGNs \citep[e.g. ][]{2016ApJ...820....5R}. Correlation between variations in photon index and AGN luminosity have been observed in several AGNs \citep{2013MNRAS.433..648F}. For bright AGNs, there is a positive correlation between the two parameters. However, low luminosity AGNs, with $2-10~\mathrm{keV}$ luminosities below $10^{-3}~L_{\rm Edd}$ were found to exhibit the opposite correlation \citep{2009MNRAS.399..349G, 2015MNRAS.447.1692Y}. The parameter $L_{\rm Edd}$ refers to the Eddington luminosity. 

The variability of X-ray spectral lines in AGNs is, however, less well established. The Fe K band generally shows a lesser degree of variability than other components of the spectrum \citep{2003ApJ...598..935M}. However, it has been possible to detect reverberation lags between the continuum and the Fe K$\alpha$ line \citep{2012MNRAS.422..129Z}.

The shape of spectral lines can further be used to investigate the structure and kinematics of the accretion disk. For instance, the shape, and width of the line are indicative of the inner radius of the emitting region. The centroid energy of a spectral line can also reflect bulk motion properties. For example, Doppler shifted emission and absorption lines can indicate the presence of outflows from the disk \citep{2017FrASS...4...16M, 2021ApJ...914...62W, 2021MNRAS.503.1442M}.

The X-ray spectrum of Cen A has been observed, and studied over five decades with many different telescopes. Over this interval, the main spectral shape has not been observed to strongly vary, despite significant luminosity variability \citep{2011ApJ...733...23R}. In this entire interval, the power law photon index has been found at $\Gamma \approx 1.8$ \citep{1978QJRAS..19....1C, 2003ApJ...593..160G, 2006ApJ...641..801R, 2011ApJ...733...23R, 2016ApJ...819..150F}. \citet{2011ApJ...733...23R} analysed \emph{RXTE} observations in the interval from 1996 to 2009, and found a consistent $\Gamma=1.822\pm0.004$. However, some spectra were also observed to be shallower than this. For instance, \citet{1978ApJ...220..790M} found $\Gamma=1.66\pm0.03$, and \citet{1981ApJ...244..429B} observed it to vary from $1.68\pm0.03$ to $1.62\pm0.03$ over the course of six months.

The power law component extends unbroken to high energies, with a consistent slope. However, there is disagreement regarding the energy at which a break to a steeper power law occurs. \citet{1981ApJ...244..429B} detected a break to $\Gamma\approx 2.0 \pm0.2$ at an energy of $E_{\rm break} = 140 ~ \mathrm{keV}$, using broadband observations from \emph{HEAO 1}. In contrast, \citet{1995ApJ...449..105K} found a break between $300-700~\mathrm{keV}$ with \emph{CGRO}. A break of $\approx 180~\mathrm{keV}$ was found by \citet{1996PASJ...48..801M} with \emph{Ginga}. \citet{2003ApJ...593..160G} fitted \emph{Beppo-SAX} data, and estimated a folding energy of $E_{\rm fold} \approx 600 ~\mathrm{keV}$. \citet{2006ApJ...641..801R, 2011ApJ...733...23R} found a lower limit for the break in the power law of $E_{\rm break} > 2 ~ \mathrm{MeV}$. A \emph{NuSTAR} spectrum provided a lower limit of: $E_{\rm fold} >1 ~ \mathrm{MeV}$ \citep{2016ApJ...819..150F}. 

The $\gamma$-ray emission from Cen A has also been detected, and studied with \emph{Fermi} and \emph{H.E.S.S} \citep{2009ApJ...695L..40A, 2010ApJ...719.1433A, 2018A&A...619A..71H}. The spectrum was observed to follow a steepening power law, with $\Gamma = 2.52 \pm 0.13_{\rm stat} \pm 0.20_{\rm sys}$. No significant variability was detected over the course of eight years of observation. 

In contrast, the Hydrogen column density has been observed to vary. \citet{2006ApJ...641..801R, 2011ApJ...733...23R} found a variation between about $10-26\times10^{22}~\mathrm{cm}^{-2}$. 

The Fe K$\alpha$ line was prominently observed in all X-ray spectra since it was first detected by \citet{1978ApJ...220..790M}. There is disagreement over whether the line varies significantly. \citet{2006ApJ...641..801R, 2011ApJ...733...23R} found the flux of the line to have a consistent value of $4.55\pm0.14\times10^{-4}~\mathrm{photons}~\mathrm{cm}^{-2}~\mathrm{s}^{-1}$. In contrast, \citet{2011ApJ...743..124F} detected a $20-30\%$ variation, and \citet{2022A&A...664A..46A} found it to vary by a factor of 10. 

Most spectral analyses did not detect any significant reflection features like the Compton hump. \citet{2011ApJ...733...23R} found an upper limit on the reflection fraction of $R < 0.005$. \citet{2016ApJ...819..150F} described $R < 0.01$, and \citet{2007ApJ...665..209M} found that the spectra were best fit with $R=0$. This is contrasted by \citet{2011ApJ...743..124F}, who detected a significant Compton hump, with a reflection fraction of $R=0.19$. 

\citet{2004ApJ...612..786E} analysed two \emph{Chandra}-HETGS spectra of Cen A that were observed in 2001. Their results are mostly consistent with the results of other spectral analyses of Cen A, except that they fitted a comparatively shallow power law slope of $\Gamma = 1.64 \pm 0.07$. They detected excess X-ray emission at $\approx 2 ~\mathrm{keV}$, which they fitted by including a second power law with a different absorption, and a photon index of $\Gamma=2$ in their spectral model. Si and S K$\alpha$ lines were also detected at $1.74~\mathrm{keV}$ and $2.30~\mathrm{keV}$, respectively. They further discussed that the $20~\mathrm{eV}$ width of the Fe K$\alpha$ line indicates that it originates from a cold, neutral medium far from the SMBH. 

\citet{2022A&A...664A..46A} analysed archival non-grating \emph{Chandra} ACIS spectra of Cen A. They also estimated the radius of the region emitting the Fe K$\alpha$ line ($0.10\pm0.05~\mathrm{pc}$), as well as the dust sublimation radius ($0.04\pm0.02~\mathrm{pc}$). 

The redshift of Cen A had been independently, and consistently measured by different groups, using different methods. \citet{1978PASP...90..237G} first measured a heliocentric redshift for the entire galaxy of $1.825\pm0.017\times10^{-3}$ from optical emission and absorption lines.  \citet{1983BAAS...15..921W}, and \citet{2006AJ....131.1163S} later measured systemic velocities corresponding to redshifts of $z=1.73\times10^{-3}$, and $z=1.826\pm0.017\times10^{-3}$, respectively. \citet{1995ApJ...449..592H} and \citet{2015A&A...574A.109W} studied the kinematics of planetary nebulae in Cen A, and found $z=1.805\pm0.023\times10^{-3}$, and $z=1.798\times10^{-3}$. \citet{2007AJ....134..494W} investigated the globular clusters in Cen A, and measured a systemic velocity corresponding to a redshift of $1.821\pm0.023\times10^{-3}$. The average of the redshift measurements with known uncertainties is $1.819\pm0.010\times10^{-3}$. We will henceforth be using this value for the Cen A redshift. 

This paper is structured as follows. Section \ref{sec:obs} describes the observations that were analysed in this paper, as well as the data reduction, and methodology that was used. The spectral analysis utilizing two different models is described in Section \ref{sec:spectra}. Section \ref{sec:dscsn} discusses the results of this analysis, and investigates different interpretations of them. Finally, section \ref{sec:conc} summarises and concludes the paper. 

\section{Observations and data analysis}\label{sec:obs}

\noindent
We performed 14 observations of Cen A with the \emph{Chandra X-ray Observatory} \citep[\textit{Chandra};][]{2000SPIE.4012....2W}, using the Advanced CCD Imaging Spectrometer \citep[ACIS;][]{2003SPIE.4851...28G} optimized for spectroscopy (ACIS-S), and the High Energy Transmission Grating Spectrometer \citep[HETGS;][]{2005PASP..117.1144C}, from January to September 2022. The observations were performed with a reduced sub-array size, to further reduce pileup of the bright central source. 

\begin{figure}[h]
\resizebox{\hsize}{!}{\includegraphics{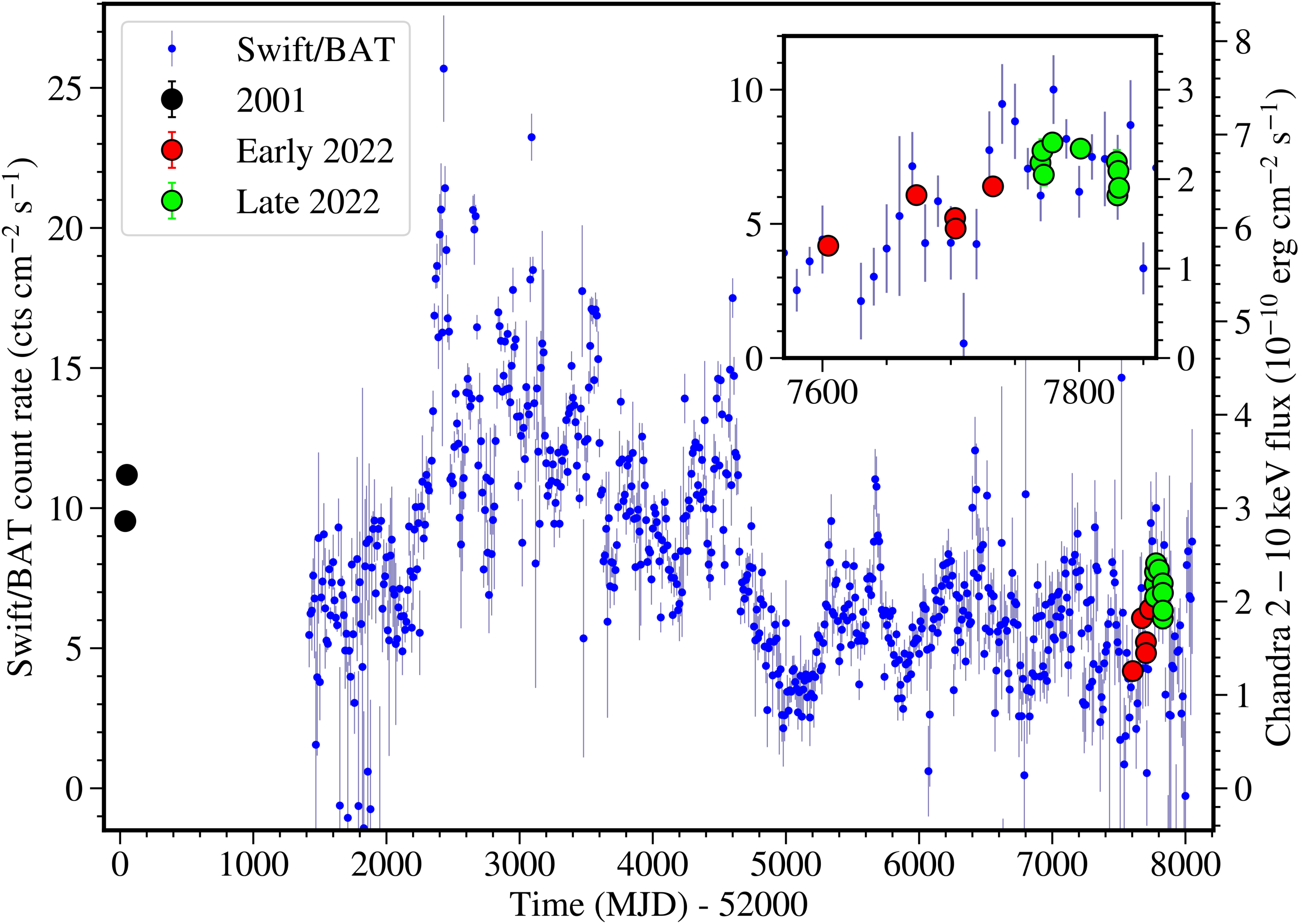}}
\caption{\emph{Swift}/BAT light curve of Cen A with overplotted $2-10 ~ \mathrm{keV}$ \emph{Chandra} fluxes for each of the observations listed in Table \ref{tab:Cobs}. The inset shows the observations obtained throughout 2022. The \emph{Swift}/BAT daily light curve was rebinned by a factor of 10, for display clarity. The colors of the \emph{Chandra} datapoints indicate the three distinct groups into which the spectra were merged. The error bars for the \emph{Chandra} fluxes are smaller than the size of the data points.  
 \label{fig:lc}}
\end{figure}

These results were compared with the two previous \emph{Chandra}-HETGS observations of Cen A that were performed in 2001, which have already been described by \citet{2004ApJ...612..786E}. These observations also used the ACIS-S, but with the full array size. Table \ref{tab:Cobs} lists the properties of all the 16 \emph{Chandra} observations analysed in this paper. The set of \emph{Chandra} datasets used can be found in the~\dataset[Chandra Data Collection (CDC) 167]{https://doi.org/10.25574/cdc.167}. 

We did not find any evidence that the different observing modes  affected any of the results discussed in the following sections. We selected consistent energy ranges to minimise the impact of the different array sizes on the spectral analysis, as will be discussed in Section \ref{sec:spectra}.

For all of these observations, we generated type-2, first order \emph{Chandra} HETGS spectra\footnote{\href{https://space.mit.edu/CXC/analysis/AGfCHRS/AGfCHRS.html}{https://space.mit.edu/CXC/analysis/AGfCHRS/AGfCHRS.html}}, using CIAO version 4.14.0, and HEASOFT version 6.30.1. First, we identified the position of the zeroth order image, and created region files for the grating source and background spectrum sky boundaries, by running \texttt{tg\_detect} with the default parameters, followed by \texttt{tg\_create\_mask}, with an HETG width factor of 18. Next, we assigned grating events to spectral orders using \texttt{tg\_resolve\_events}, with a pixel randomization half-width of 0.5. We created $+1$ and $-1$ grating order type II PHA spectral files for the source and background using \texttt{tgextract}, with the default parameters. We generated RMF and ARF files for the spectra with \texttt{mkgrmf} and \texttt{fullgarf}, using the default parameters. Finally, we combined the positive and negative orders of individual, or groups of observations using \texttt{combine\_grating\_spectra}. We kept the HEG and MEG spectra separate.

Fig. \ref{fig:lc} shows the \emph{Swift}/BAT light curve of Cen A, with overplotted \emph{Chandra} fluxes at the times of the observations. Cen A was brighter in 2001 than in 2022, but it had been even brighter during the intervening interval.

We used XSPEC \citep{1996ASPC..101...17A} version 12.12.1 to fit all \emph{Chandra} spectra. We assumed solar abundances, as described by \citet{2000ApJ...542..914W}, and the cross-sections defined by \citet{1996ApJ...465..487V}. The spectra were not rebinned prior to fitting, but were subsequently rebinned for visual clarity in the following figures. The best fits were found by minimising the C-statistic \citep{1979ApJ...228..939C}. 

\begin{table}[ht]
\centering
\setlength{\tabcolsep}{3pt}
\def\arraystretch{1.0}
\begin{tabular}{l|l|l|l|l}
    \textbf{ObsID} & \textbf{Date} & $\boldsymbol{T} (\mathrm{ks})$ & $\boldsymbol{F_{2-10}}$ & $\boldsymbol{C} (10^3)$ \\ \hline
    1600 & 2001-05-09 & 46.85 & $2.96\pm0.01$ & $37.724$ \\ 
    1601 & 2001-05-21 & 51.51 & $3.43\pm0.01$ & $48.047$ \\ \hline
    24322 & 2022-01-25 & 25.32 & $1.25\pm0.02$ & $8.287$ \\ 
    23823 & 2022-04-04 & 27.81 & $1.87\pm0.03$ & $13.462$ \\ 
    24321 & 2022-05-04 & 13.95 & $1.56\pm0.02$ & $5.879$ \\ 
    26405 & 2022-05-04 & 13.95 & $1.41\pm0.03$ & $5.332$ \\ 
    24319 & 2022-06-02 & 28.27 & $1.97\pm0.03$ & $14.576$ \\ \hline
    24325 & 2022-07-09 & 29.66 & $2.25\pm0.01$ & $17.477$ \\ 
    24323 & 2022-07-11 & 20.51 & $2.38\pm0.02$ & $12.509$ \\ 
    26453 & 2022-07-12 & 9.33 & $1.92\pm0.04$ & $5.021$ \\ 
    24318 & 2022-07-19 & 28.79 & $2.49\pm0.03$ & $18.053$ \\
    24324 & 2022-08-10 & 29.66 & $2.40\pm0.03$ & $18.097$ \\ 
    24320 & 2022-09-07 & 13.02 & $2.19\pm0.04$ & $7.181$ \\ 
    24326 & 2022-09-07 & 13.02 & $1.78\pm0.04$ & $6.378$ \\ 
    27344 & 2022-09-08 & 13.02 & $2.07\pm0.04$ & $6.919$ \\ 
    27345 & 2022-09-09 & 13.02 & $1.91\pm0.04$ & $6.757$ \\ 
\end{tabular}
\caption{The list of \emph{Chandra} observations of Cen A that we used. $T$ denotes the total exposure time of each observation, and $F_{2-10}$ refers to the absorbed $2-10~\mathrm{keV}$ HEG flux, and is listed in units of $10^{-10}~\mathrm{erg} ~ \mathrm{cm}^{-2} ~ \mathrm{s}$. The parameter $C$ refers to the sum of the total number of source counts of the MEG from $1.65-3.5~\mathrm{keV}$, and the HEG from $3.5-10.0~\mathrm{keV}$. The horizontal lines distinguish between the 2001, early 2022, and late 2022 groups of observations.
\label{tab:Cobs}}
\end{table}

\section{Spectral analysis}\label{sec:spectra}

\noindent
Initially, we fitted the spectra of individual observations. As they have a limited sensitivity, we fitted them with a comparatively simple phenomenological model that only described the three most important features of the spectrum; the power law, the Fe K$\alpha$ line, and the Hydrogen absorption. In XSPEC notation, the model we used is: \texttt{constant*ztbabs*(powerlaw+gauss)}. 

The HETGS MEG and HEG spectra were fitted jointly for each observation. The \texttt{constant} component was used to account for slight normalization differences between them, with a fixed value of $1.0$ for the HEG spectrum. The normalizations of the HEG and MEG spectra differed by at most $7\%$ in individual observations. This agrees well with previous results that found approximately $8\%$ difference between the spectral normalizations of HEG and MEG\footnote{\href{https://space.mit.edu/ASC/calib/heg_meg/}{https://space.mit.edu/ASC/calib/heg\_meg/}}. The \texttt{ztbabs} component describes the total absorption, featuring contributions from Cen A and the Milky Way. Since we expect most of the absorption to take place in Cen A, we set the redshift in \texttt{ztbabs} to the value for the host galaxy.  The component \texttt{gauss} is used to describe the Fe K$\alpha$ line, on top of the \texttt{powerlaw} continuum. Other emission or absorption lines were not resolved in the spectra of most of the individual observations. 

Table \ref{tab:Cobs} also lists the $2-10 ~ \mathrm{keV}$ HEG fluxes measured in each of the observations. This absorbed continuum flux varied by a factor of $2.74\pm0.05$, between $1.25\pm0.02\times10^{-10}~\mathrm{erg} ~ \mathrm{cm}^{-2} ~ \mathrm{s}^{-1}$ and $3.43\pm0.01\times10^{-10}~\mathrm{erg} ~ \mathrm{cm}^{-2} ~ \mathrm{s}^{-1}$. The corresponding unabsorbed fluxes range from $2.33\pm0.05\times10^{-10}~\mathrm{erg} ~ \mathrm{cm}^{-2} ~ \mathrm{s}^{-1}$ to $5.97\pm0.02\times10^{-10}~\mathrm{erg} ~ \mathrm{cm}^{-2} ~ \mathrm{s}^{-1}$. \citet{2009MNRAS.394..660C} measured the Cen A SMBH to have a mass of $5.5\pm3.0\times10^7 ~\mathrm{M}_{\odot}$, with the error quoted at the $3\sigma$ level. As we are using $1\sigma$ errors throughout this paper, we will assume that the $1\sigma$ errors of the mass are $\pm1.0\times10^7 ~\mathrm{M}_{\odot}$. Using this mass, the measured distance of $3.8\pm0.1~\mathrm{Mpc}$, and the assumption of an isotropic flux, we can calculate the source luminosity to range from $4.43\pm0.16\times10^{41}~\mathrm{erg} ~\mathrm{s}^{-1}= 6.4\pm1.2 \times 10^{-5}~ L_{\rm Edd}$, to $1.03\pm0.03\times10^{42}~\mathrm{erg} ~\mathrm{s}^{-1}= 1.49\pm0.27 \times 10^{-4} ~ L_{\rm Edd}$, respectively. These luminosities do not include a bolometric correction.

Individual spectra lacked the sensitivity to constrain various spectral parameters, investigate weaker features, and fit more complex physical models. We found that the main difference between the spectral fits of individual observations was the normalization of the power law. This caused most of the variation in the $2-10~\mathrm{keV}$ flux shown in Table \ref{tab:Cobs}. Most other spectral parameters had consistent values for spectra of individual observations. Furthermore, we did not detect any indication of significant variation in the spectra obtained within a few weeks or even months of each other. 

Therefore, to better investigate the Cen A spectra, we merged the data from all observations into one of three groups, corresponding to the two observations in 2001, the first five observations in 2022, and the subsequent nine observations in the same year. The observations in 2022 were divided into two groups to investigate changes occurring on timescales of a few months. The selection of observations to include in the early and late 2022 grouped spectra was based on the close proximity in time of the final nine observations, as well as the lower flux detected in most of the earlier five observations (see Fig. \ref{fig:lc}).

\begin{figure}[h]
\resizebox{\hsize}{!}
{\includegraphics{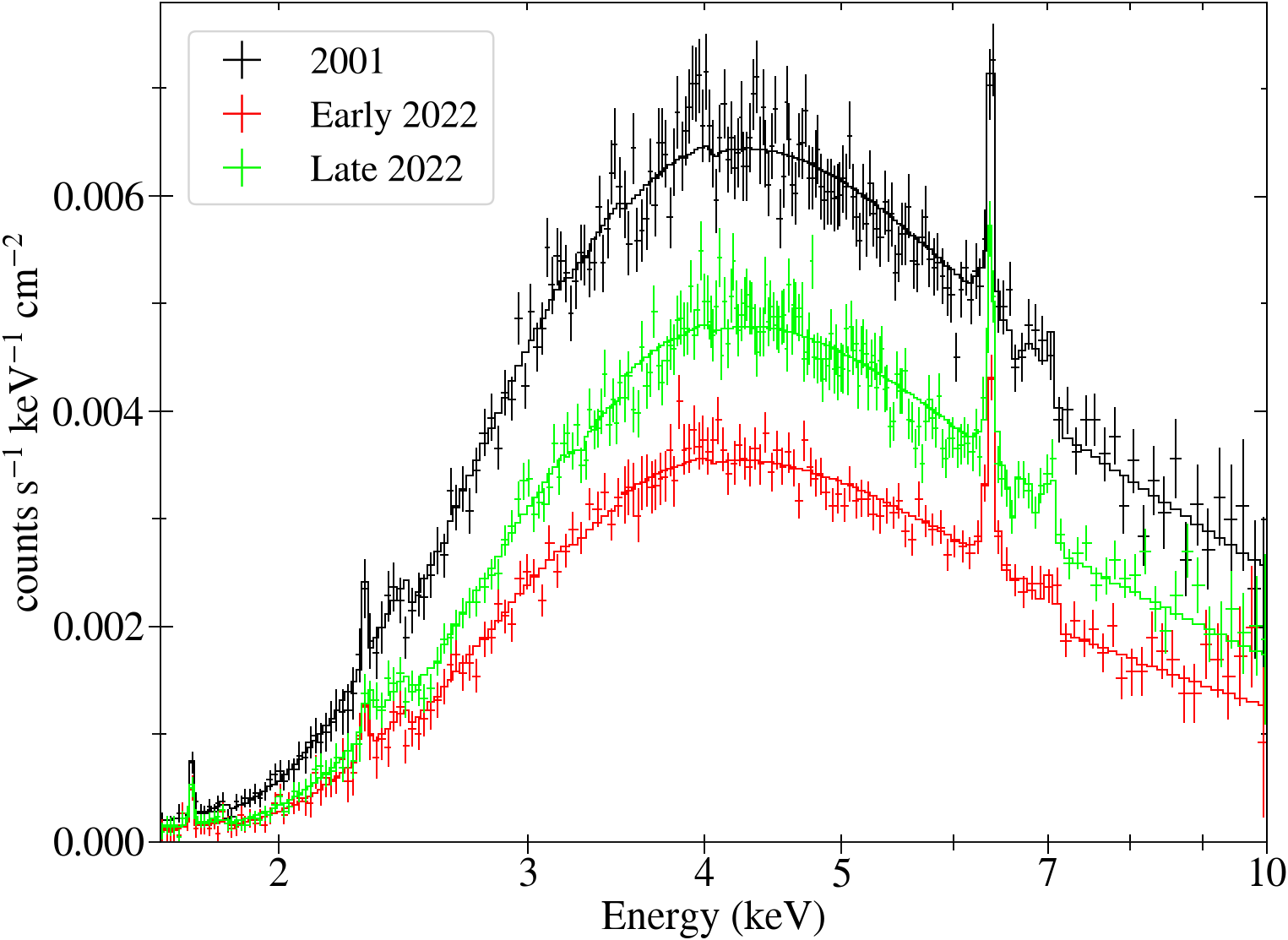}}
\caption{$2-10~\mathrm{keV}$ spectra of the grouped 2001, early 2022, and late 2022 spectra. This figure contains both HEG and MEG spectra, which are depicted with the same color, unlike the following spectra. It also shows the best fit to the data, using model B. The spectra have been rebinned for visual clarity. 
 \label{fig:specall}}
\end{figure}

To investigate the differences occurring between 2001 and 2022 as best as possible, we also created a spectrum that combined all observations from 2022. Finally, to obtain the best constraints on some spectral properties that were found to be consistent over this 21-year interval, we also analyzed the spectrum created by grouping all data together. We will subsequently refer to these five groups of merged spectra as the 2001, early 2022, late 2022, 2022, and total spectra. The total exposure time of these five groups of observations is: $98.36~\mathrm{ks}$, $109.30~\mathrm{ks}$, $170.03~\mathrm{ks}$, $279.33~\mathrm{ks}$, and $377.69~\mathrm{ks}$, respectively. The total number of counts from $1.65-3.5~\mathrm{keV}$ in the MEG, and $3.5-10.0~\mathrm{keV}$ in the HEG is: $85771$, $47536$, $98392$, $145928$, and $231699$, respectively. The three non-overlapping groups of observations are indicated via different colors in Fig. \ref{fig:lc}. Their merged spectra are shown in Fig. \ref{fig:specall}, indicating their qualitatively similar spectral shapes. This is essential for fitting the total spectrum. 

The background spectrum is mostly flat across the energy range for both the MEG and HEG. It is consistently at least one order of magnitude below the source spectrum between $2-10~\mathrm{keV}$. Even at the lowest energy we investigated, of $1.65~\mathrm{keV}$, the background spectrum is still fainter than the source spectrum by a factor of at least 3.

The part of the spectrum below $2.5~\mathrm{keV}$ is brighter than expected from extrapolating the higher energy spectral shape. This feature was first described by \citet{1997ApJ...475..118T}, and studied in greater detail in the 2001 \emph{Chandra} observations by \citet{2004ApJ...612..786E}. To accurately describe it, we included a leaked power law component that is only weakly absorbed. This is still a nuclear emission which is either the result of a leaky absorber or emission from the innermost part of the jet. Therefore, all of the following spectral fits include the additional terms \texttt{constant*tbabs*powerlaw}. The \texttt{constant} was set to a low initial value, in order to only describe the slight discrepancies observed at low energies, rather than the main spectral shape. The Hydrogen column density in \texttt{tbabs} was set equal to the weighted average of the Milky Way absorption in the direction of Cen A, of $N_{\rm H} = 2.35\times10^{20}~\mathrm{cm}^{-2}$\citep{2016A&A...594A.116H}. Finally, the slope and normalization of the \texttt{powerlaw} component were set equal to those of the power law describing the main part of the spectrum. 

Nevertheless, the MEG spectrum is too faint to accurately describe below about $1.65~\mathrm{keV}$. At lower energies, the source is also comparable to the background level. Therefore, we only investigated the spectra at greater energies. The smaller size of the sub-array used for the 2022 observations limits the HEG energy range to above about $2.6~\mathrm{keV}$\footnote{\href{https://cxc.harvard.edu/proposer/POG/html/chap8.html}{https://cxc.harvard.edu/proposer/POG/html/chap8.html}}. Furthermore, the $+1$ and $-1$ arms of all the HEG spectra significantly deviated at about $3.3~\mathrm{keV}$, with the appearance of an apparent emission line in the -1 arm. This was caused by a large drop in the response efficiency, possibly due to a chip gap. Above $3.5~\mathrm{keV}$, the HEG spectra have a greater sensitivity than the MEG spectra, but the two are consistent with each other, when correcting for their slightly different normalizations. Therefore, we selected an energy range of $1.65-3.5~\mathrm{keV}$ for the co-added MEG spectra, and $3.5-10.0~\mathrm{keV}$ for the co-added HEG spectra. 

The low energy part of the spectra also exhibit several peaks away from the continuum shape. \citet{2004ApJ...612..786E} detected Si and S K$\alpha$ lines, which have rest frame energies of $1.740~\mathrm{keV}$, and $2.307~\mathrm{keV}$, respectively. In the merged spectrum of all observations, we saw other features that might be interpreted as emission lines.

In order to determine the statistical significance of these lines, we fitted a segment of the total spectrum in the energy range around each line with both \texttt{ztbabs*powerlaw}, and \texttt{ztbabs*(powerlaw+gauss)}. We calculated the Bayesian Information Criterion \citep[BIC;][]{1978AnSta...6..461S} of the two fits, and only selected lines as statistically significant if the addition of the \texttt{gauss} component resulted in a lower BIC. Of all the deviations from the spectrum that we observed, only the Si and S K$\alpha$ lines satisfied this condition, so we will subsequently only describe them. 

We fitted the five groups of \emph{Chandra} spectra with two main types of models, that each include the features described above. Figs. \ref{fig:specmod01}, \ref{fig:specmod22}, as well as Figs. \ref{fig:specmod22el} and \ref{fig:specmodall} in Appendix \ref{sec:app1} show the best fits to the five spectra, using models A and B.

\subsection{Model A}

\begin{figure}[htp]
\resizebox{\hsize}{!}{\includegraphics{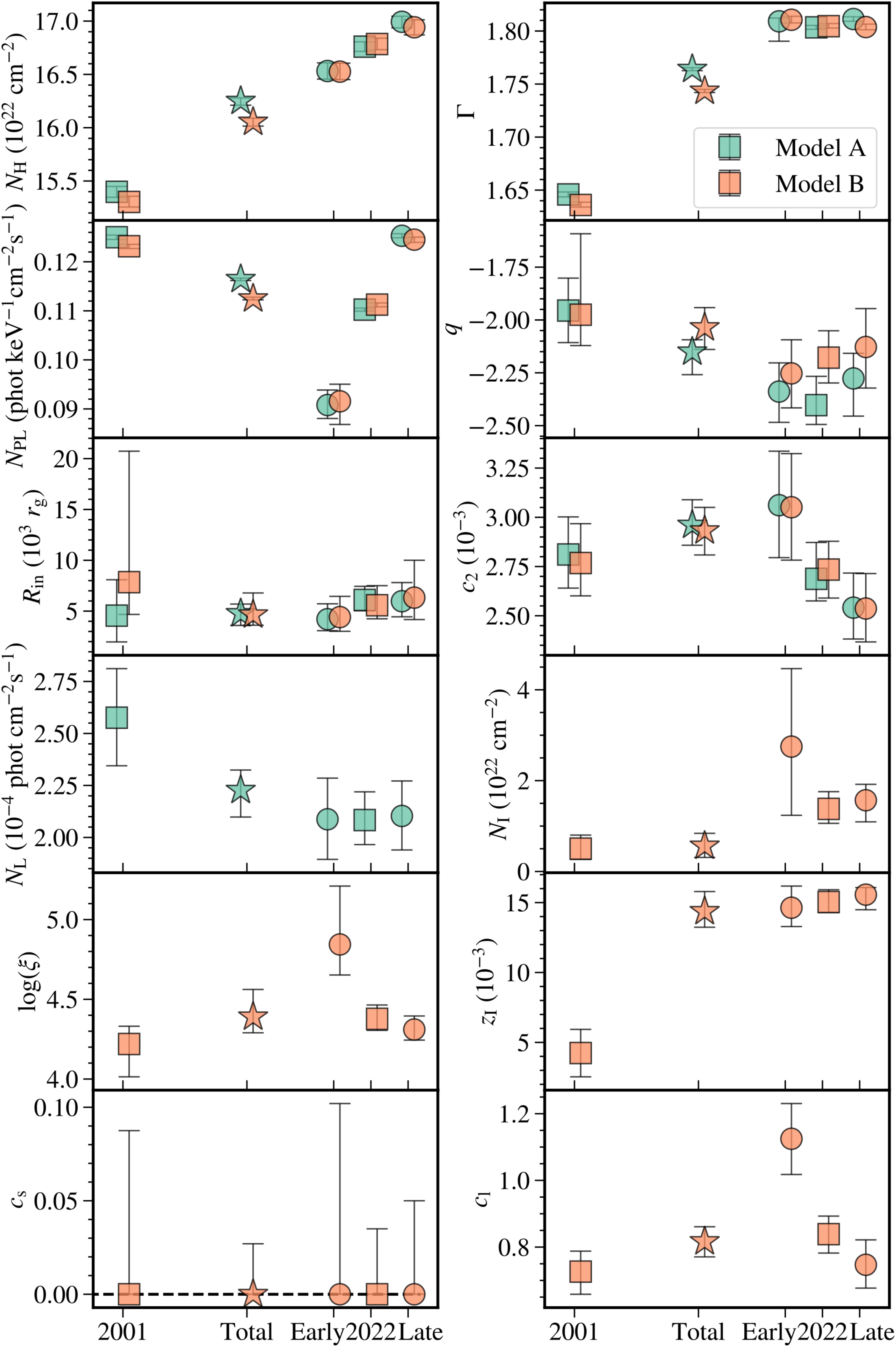}}
\caption{Variation of the best fit parameters of models A and B between the 2001, early 2022, late 2022, 2022, and total spectra. The value of the best fit of a parameter in the total spectrum is unreliable, if it varied significantly from 2001 to 2022. 
 \label{fig:parcompAB}}
\end{figure}

\noindent
Model A is a phenomenological model describing the spectrum as a set of absorbed power laws, absorption features, and emission lines. In XSPEC notation, it is written as: \texttt{constant\textsubscript{1}*(tbabs*ztbabs*(powerlaw+diskline+ gauss\textsubscript{1}+gauss\textsubscript{2})+constant\textsubscript{2}*tbabs*powerlaw)}. The parameter \texttt{constant\textsubscript{1}} describes slight normalization differences between the HEG and MEG spectra. The Milky Way absorption is parameterized by \texttt{tbabs}, with a fixed column density of $N_{\rm H} = 2.35\times10^{20}~\mathrm{cm}^{-2}$\citep{2016A&A...594A.116H}. The Cen A absorption in the line of sight is described by \texttt{ztbabs}, with a redshift of $1.819\times10^{-3}$, and a column density that is free to vary. The \texttt{diskline} model is a physically accurate description of an emission line profile in an accretion disk, which we used to fit the Fe K$\alpha$ line. The two \texttt{gauss} components represent the emission lines of Si, and S K$\alpha$. The parameters of components with identical names were linked. Components with different numbers describe different features, so were not linked. Table \ref{tab:modA1} lists the best fitting parameters for model A applied to the five groups of spectra, as well as their $2-10~\mathrm{keV}$ fluxes. The results are also depicted in Fig. \ref{fig:parcompAB}. 

The merged 2001 spectrum was the brightest, and featured a lower Hydrogen column density and power law index than both the early and late spectra from 2022. The $2-10~\mathrm{keV}$ flux dropped by a factor of $1.89\pm0.01$ from 2001 to early 2022. Meanwhile, the Hydrogen column density increased by $1.59^{+0.07}_{-0.08} \times 10^{22} ~\mathrm{cm}^{-2}$ from 2001 to late 2022. The fits also found a slight difference of $0.46^{+0.09}_{-0.10}\times 10^{22} ~\mathrm{cm}^{-2}$ between the column density for the early and late 2022 spectra. The photon index of the power law also increased by $0.165\pm0.003$ from 2001 to late 2022. Even though the increase in the Hydrogen column density and the power law index also contributed to the reduction in flux, that was predominantly caused by a decrease in the power law normalization.

There is a degeneracy between these two parameters, so we sought to determine whether the observed variation in the photon index could instead be attributed to variations in the Hydrogen column density, and other fit parameters. For this purpose, we fitted all five spectra again with model A, but with the photon index set to the fixed value of $\Gamma = 1.815$. That is the photon index found by \citet{2016ApJ...819..150F} for a \emph{NuSTAR} spectrum, which also agrees well with the results of several other X-ray spectral analyses of Cen A \citep{1978QJRAS..19....1C, 2003ApJ...593..160G, 2006ApJ...641..801R, 2011ApJ...733...23R}. 

Comparing the results of the two spectral fits, we found that the three merged spectra that were generated from the 2022 observations all had a lower BIC when the photon index was fixed at $\Gamma=1.815$. The reason for this, is that fits with a free $\Gamma$ already found it to be close to $1.815$. However, the best fit photon indices for the late 2022, and entire 2022 spectra with a free $\Gamma$ are inconsistent with a value of $1.815$, within $1\sigma$ errors. This indicates that the errors in the parameter may be underestimated.

In contrast, the BIC increased by $35.36$ for the 2001 spectrum when freezing the photon index at $\Gamma=1.815$. This demonstrates that the best fit value for it, of $1.646\pm0.002$ is indeed inconsistent with $\Gamma=1.815$. Therefore, the photon index did significantly change from 2001 to 2022. The variation of $N_{\rm H}$ and $\Gamma$ between 2001 and 2022 cause the best fit values for these two parameters in the total spectrum to be unreliable.

\begin{figure}[htp]
\resizebox{\hsize}{!}{\includegraphics{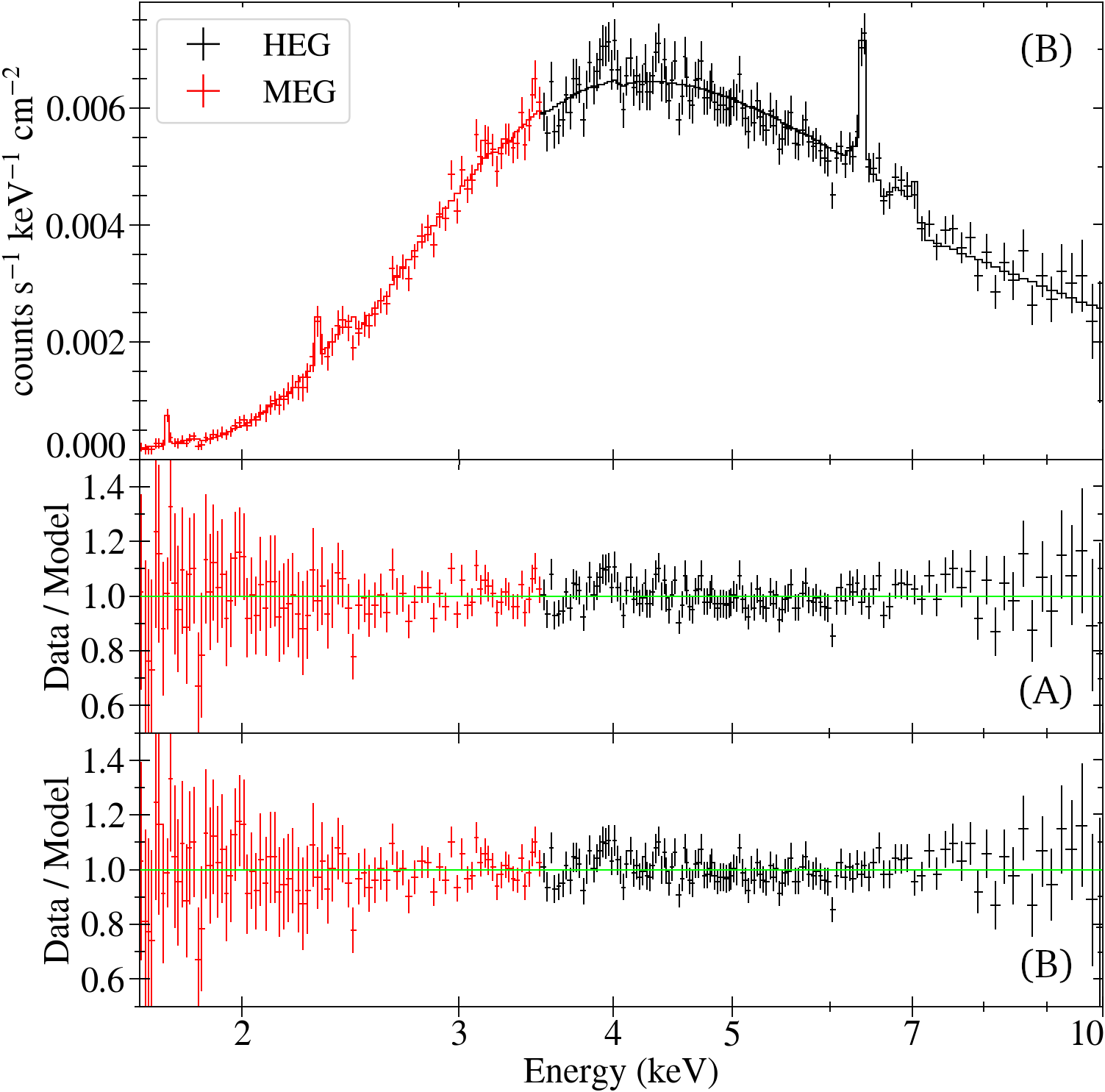}}
\caption{The best fit spectra, and the ratio of the data to the folded model, for the grouped 2001 spectrum. The first panel shows the spectrum, and the best fit using model B. The subsequent two panels depict the residuals normalized by the folded model, for spectral models A and B. The spectra were rebinned for visual clarity. 
 \label{fig:specmod01}}
\end{figure}

The leaked power law component has a variable strength, but nevertheless remains close to $c_2\approx 2.8\times10^{-3}$ in all spectra. It can vary on short timescales, as it was found to decrease from $3.06\pm0.27\times10^{-3}$ to $2.54^{+0.18}_{-0.16}\times10^{-3}$ from early to late 2022. This may, however, be due to a degeneracy between the leaked power law strength and the Hydrogen column density.

\begin{table*}[ht]
\centering
\setlength{\tabcolsep}{3pt}
\def\arraystretch{1.1}
\begin{tabular}{l l l || l | l l l | l}
    \textbf{Component} & & \textbf{Units} & \textbf{2001} & \textbf{Early 2022} & \textbf{Late 2022} & \textbf{2022} & \textbf{Total}\\ \hline \hline
    \texttt{phabs} & $\boldsymbol{N_{\rm H}}$ & $10^{22}~\mathrm{cm}^{-2}$ & $15.40\pm0.05$ & $16.53\pm0.08$ & $16.99^{+0.05}_{-0.06}$ & $16.76\pm0.04$ & $16.25^{+0.03}_{-0.04}$ \\ \hline
    \texttt{powerlaw} & $\boldsymbol{\Gamma}$ & & $1.646\pm0.002$ & $1.809^{+0.003}_{-0.019}$ & $1.811\pm0.002$ & $1.803\pm0.002$ & $1.764\pm0.001$ \\ 
    & $\boldsymbol{N_{\rm PL}}$ & $\mathrm{photons}~\mathrm{keV}^{-1}~\mathrm{cm}^{-2}~ \mathrm{s}^{-1}$ & $0.1250\pm0.0004$ & $0.0908^{+0.0031}_{-0.0027}$ & $0.1253\pm0.0004$ & $0.1102\pm0.0003$ & $0.1164^{+0.0002}_{-0.0003}$ \\ \hline
    \texttt{diskline} & $\boldsymbol{E_{\rm l}}$ & $\mathrm{keV}$ & $6.382^{+0.004}_{-0.002}$ & $6.378\pm0.004$ & $6.379^{+0.004}_{-0.002}$  & $6.378^{+0.003}_{-0.002}$ & $6.381\pm0.002$ \\ 
    Fe K$\alpha$ & $\boldsymbol{z}$ & $10^3$ & $2.75^{+0.33}_{-0.61}$ & $3.37\pm0.68$ & $3.26^{+0.39}_{-0.63}$ & $3.37^{+0.39}_{-0.44}$ & $2.95^{+0.28}_{-0.31}$ \\ 
    & $\boldsymbol{q}$ & & $-1.96\pm0.15$ & $-2.34^{+0.14}_{-0.15}$ & $-2.28^{+0.12}_{-0.18}$  & $-2.40^{+0.14}_{-0.09}$ & $-2.15^{+0.06}_{-0.11}$ \\
    & $\boldsymbol{R_{\rm in}}$ & $10^3 ~ r_{\rm g}$ & $4.6^{+3.5}_{-2.6}$ & $4.2^{+1.5}_{-1.1}$ & $6.0^{+1.8}_{-1.5}$  & $6.1^{+1.3}_{-1.1}$ & $4.8^{+0.9}_{-1.2}$ \\ 
    & $\boldsymbol{N_{\rm L}}$ & $10^{-4} ~ \mathrm{photons}~\mathrm{cm}^{-2}~ \mathrm{s}^{-1}$ & $2.58\pm0.23$ & $2.09\pm0.20$ & $2.10\pm0.17$ & $2.08^{+0.14}_{-0.12}$ & $2.23^{+0.1}_{-0.13}$ \\ 
    & $\boldsymbol{F_{\rm L}}$ & $10^{-12} ~ \mathrm{erg}~\mathrm{cm}^{-2}~ \mathrm{s}^{-1}$ & $2.63\pm0.24$ & $2.13\pm0.20$ & $2.15\pm0.17$ & $2.14\pm0.13$ & $2.26\pm0.11$ \\
    & $\boldsymbol{EW_{\rm L}}$ & $\mathrm{eV}$ & $43.4\pm3.9$ & $63.5\pm6.3$ & $48.1\pm3.8$ & $53.1\pm3.2$ & $50.1\pm2.5$ \\ \hline
    \texttt{gauss\textsubscript{1}} & $\boldsymbol{E_{\rm l1}}$ & $\mathrm{keV}$ & $1.7371^{+0.0006}_{-0.0007}$ & $1.736\pm0.001$ & $1.737\pm0.001$ & $1.7362^{+0.0009}_{-0.0005}$ & $1.7368^{+0.0004}_{-0.0007}$ \\ 
    Si K$\alpha$ & $\boldsymbol{z_2}$ & $10^{-3}$ & $1.52^{+0.41}_{-0.35}$ & $2.15^{+0.45}_{-0.60}$ & $1.82^{+0.59}_{-0.55}$ & $2.08^{+0.30}_{-0.52}$ & $1.72^{+0.37}_{-0.22}$ \\
    & $\boldsymbol{\sigma_1}$ & $\mathrm{eV}$ & $3.04^{+0.79}_{-0.89}$ & $2.9^{+1.0}_{-0.8}$ & $4.7^{+1.2}_{-1.0}$ & $3.94^{+0.82}_{-0.67}$ & $3.71^{+0.47}_{-0.64}$ \\
    & $\boldsymbol{N_{\rm L1}}$ & $10^{-4} ~ \mathrm{photons}~\mathrm{cm}^{-2}~ \mathrm{s}^{-1}$ & $33.9\pm4.7$ & $40.4^{+8.5}_{-7.8}$ & $62.0^{+9.7}_{-9.0}$ & $51.0^{+6.6}_{-5.8}$ & $44.7\pm4.1$ \\ 
    & $\boldsymbol{F_{\rm L1}}$ & $10^{-12} ~ \mathrm{erg}~\mathrm{cm}^{-2}~ \mathrm{s}^{-1}$ & $9.4\pm1.3$ & $11.2^{+2.4}_{-2.2}$ & $17.3\pm2.6$ & $14.3\pm1.7$ & $12.4\pm1.1$ \\ \hline    
    \texttt{gauss\textsubscript{2}} & $\boldsymbol{E_{\rm l2}}$ & $\mathrm{keV}$ & $2.304^{+0.002}_{-0.001}$ & $2.300\pm0.005$ & $2.313\pm0.006$ & $2.313\pm0.006$ & $2.304\pm0.002$ \\ 
    S K$\alpha$ & $\boldsymbol{z_2}$ & $10^{-3}$ & $1.72^{+0.60}_{-0.68}$ & $3.2^{+2.2}_{-1.9}$ & $-2.3\pm2.8$ & $0.6^{+1.8}_{-1.6}$ & $1.30\pm0.78$ \\
    & $\boldsymbol{\sigma_2}$ & $\mathrm{eV}$ & $2.1^{+2.2}_{-2.1}$ & $12.2^{+4.3}_{-2.9}$ & $14.7^{+6.8}_{-4.0}$ & $11.3^{+3.9}_{-3.0}$ & $6.5^{+2.6}_{-2.0}$ \\
    & $\boldsymbol{N_{\rm L2}}$ & $10^{-4} ~ \mathrm{photons}~\mathrm{cm}^{-2}~ \mathrm{s}^{-1}$ & $3.23^{+0.9}_{-0.77}$ & $4.2^{+1.3}_{-1.2}$ & $4.1^{+1.3}_{-1.2}$ & $3.55^{+0.92}_{-0.74}$ & $3.36^{+0.41}_{-0.78}$ \\ 
    & $\boldsymbol{F_{\rm L2}}$ & $10^{-12} ~ \mathrm{erg}~\mathrm{cm}^{-2}~ \mathrm{s}^{-1}$ & $1.20\pm0.31$ & $1.55\pm0.47$ & $1.54\pm0.47$ & $1.34\pm0.30$ & $1.17\pm0.21$ \\ \hline
    \texttt{constant\textsubscript{2}} & $\boldsymbol{c_2}$ & $10^{-3}$ & $2.81^{+0.19}_{-0.17}$ & $3.06\pm0.27$ & $2.54^{+0.18}_{-0.16}$ & $2.69^{+0.18}_{-0.11}$ & $2.96^{+0.13}_{-0.10}$ \\ \hline \hline
    & $\boldsymbol{F_{2-10}}$ & $10^{-10} ~ \mathrm{erg}~\mathrm{cm}^{-2}~ \mathrm{s}^{-1}$ & $3.20\pm0.01$ & $1.689\pm0.008$ & $2.286\pm0.007$ & $2.054\pm0.005$ & $2.363\pm0.005$  \\ \hline \hline
    & $\boldsymbol{C}$ & & $1781.63$ & $1817.27$ & $1816.20$ & $1895.73$ & $1927.01$ \\
    & $\boldsymbol{BIC}$ & & $1893.32$ & $1928.96$ & $1927.89$ & $2007.42$ & $2038.70$ \\
     
\end{tabular}
\caption{Best fit properties of the fits to the spectra with XSPEC model A, \texttt{constant\textsubscript{1}*(tbabs*ztbabs*(powerlaw+diskline+ gauss\textsubscript{1}+gauss\textsubscript{2})+constant\textsubscript{2}*tbabs*powerlaw)}. The parameter $N_{\rm H}$ describes the Hydrogen column density. The power law component is parameterized by the index ($\Gamma$) and its normalization ($N_{\rm PL}$). The \texttt{diskline} model describing the Fe K$\alpha$ line has an energy of $E_{\rm l}$, an emissivity power law index of $q$, an inner accretion disk radius of $R_{\rm in}$, and a normalization of $N_{\rm L}$. The flux of the line is denoted as $F_l$, and its equivalent width is $EW_{\rm L}$. The two other emission lines are described by Gaussian functions with a standard deviation of $\sigma$. The redshift, $z$, describes the shift of the centroid of the emission line relative to its rest frame energy. The strength of the leaked component of the power law is described by the constant, $c_2$. $F_{2-10}$ denotes the absorbed $2-10~\mathrm{keV}$ flux of the spectrum. The Cash statistic ($C$), and its corresponding BIC are listed at the bottom of the table. For all of these fits, there are 1713 bins, and 1699 degrees of freedom.
\label{tab:modA1}}
\end{table*}

As Fig. \ref{fig:specall} shows, the Fe K$\alpha$ line is narrow, which means that either the inclination of the disk is low, or its inner radius is large, or both. The inclination of the AGN is not known. Although \citet{2007ApJ...671.1329N} found a mean inclination of the warped gas disk of $\approx34\degree$, inconsistent values have been found for the inclination of the jet. For instance, \citet{1979AJ.....84..284D} found an inclination of $72\pm3\degree$, and \citet{1994ApJ...426L..23S} measured it to be $61\pm5\degree$. However, \citet{2003ApJ...593..169H} argued for an inclination of $\approx 15\degree$, and \citet{2014A&A...569A.115M} found it to be $12-45\degree$. 

A too small inclination is at odds with the identification of Cen A as a Seyfert 2 galaxy \citep{2011A&A...531A..70B}, given the AGN unification model \citep{1993ARA&A..31..473A}. Furthermore, the galaxy has a high inclination, and both the jet and counterjet from the AGN are visible, even though the counterjet appears noticeably weaker. 

We investigated whether \emph{Chandra} spectra could distinguish between different inclinations, as the shape of the emission line described by \texttt{diskline} is inclination dependent. For this purpose, we jointly fitted the spectra of all individual observations simultaneously with model A. The inclination in each of the fits was fixed to a particular value between $10\degree$ and $90\degree$, in steps of $10\degree$. We allowed the inner radius and emissivity to vary. In all cases, comparably good spectral fits were found, so we concluded that these \emph{Chandra} spectra were insensitive to different inclinations.

An inclination of $60\degree$ was assumed in previous studies of the Cen A X-ray spectrum \citep{2011ApJ...743..124F, 2016ApJ...819..150F}, as it is equally likely to find a higher or a lower inclination, given a uniform, isotropic distribution of angles. As we were unable to constrain the inclination, we will also assume it to be $60\degree$. 

This choice of inclination requires a large inner and outer radius of the disk for an accurate description of the shape of the Fe K$\alpha$ line. The fits can only constrain the outer radius to within an order of magnitude, so we set it to a value of $10^6~r_{\rm g}$, where $r_{\rm g} = GM / c^2$ is the gravitational radius. The best fit values found for the inner radius depend on the choice of inclination for the fit. At a lower inclination, comparable fits are found with smaller inner and outer disk radii. The errors quoted for the inner radii do not incorporate the uncertainty of the inclination of the system. The redshift of the Fe K$\alpha$ line is unaffected by the inclination selected for the fits.

We allowed the emissivity index, $q$, to vary freely. The resulting fits were significantly better than those obtained by freezing it at a value of $q=-3$.

We compared the measured centroid energies of the three fluorescent emission lines with their rest frame energies. The K$\alpha_1$ and K$\alpha_2$ lines are so close in energy, that they merge to form a single emission line in the observed spectra. The laboratory-measured rest frame energies are found by calculating the weighted average of the K$\alpha_1$ and K$\alpha_2$ lines, using a 2:1 ratio of intensities. The resulting Fe, Si, and S K$\alpha$ rest frame energies are $6.3996796\pm0.0000074$, $1.739788\pm0.000017$, and $2.307490\pm0.000026~\mathrm{keV}$, respectively\footnote{\href{https://physics.nist.gov/PhysRefData/XrayTrans/Html/search.html}{https://physics.nist.gov/PhysRefData/XrayTrans/Html/search.html}}. The redshifts associated with the difference between the measured centroid energy, and these rest frame energies have also been listed in Table \ref{tab:modA1}. 

\begin{figure}[htp]
\resizebox{\hsize}{!}{\includegraphics{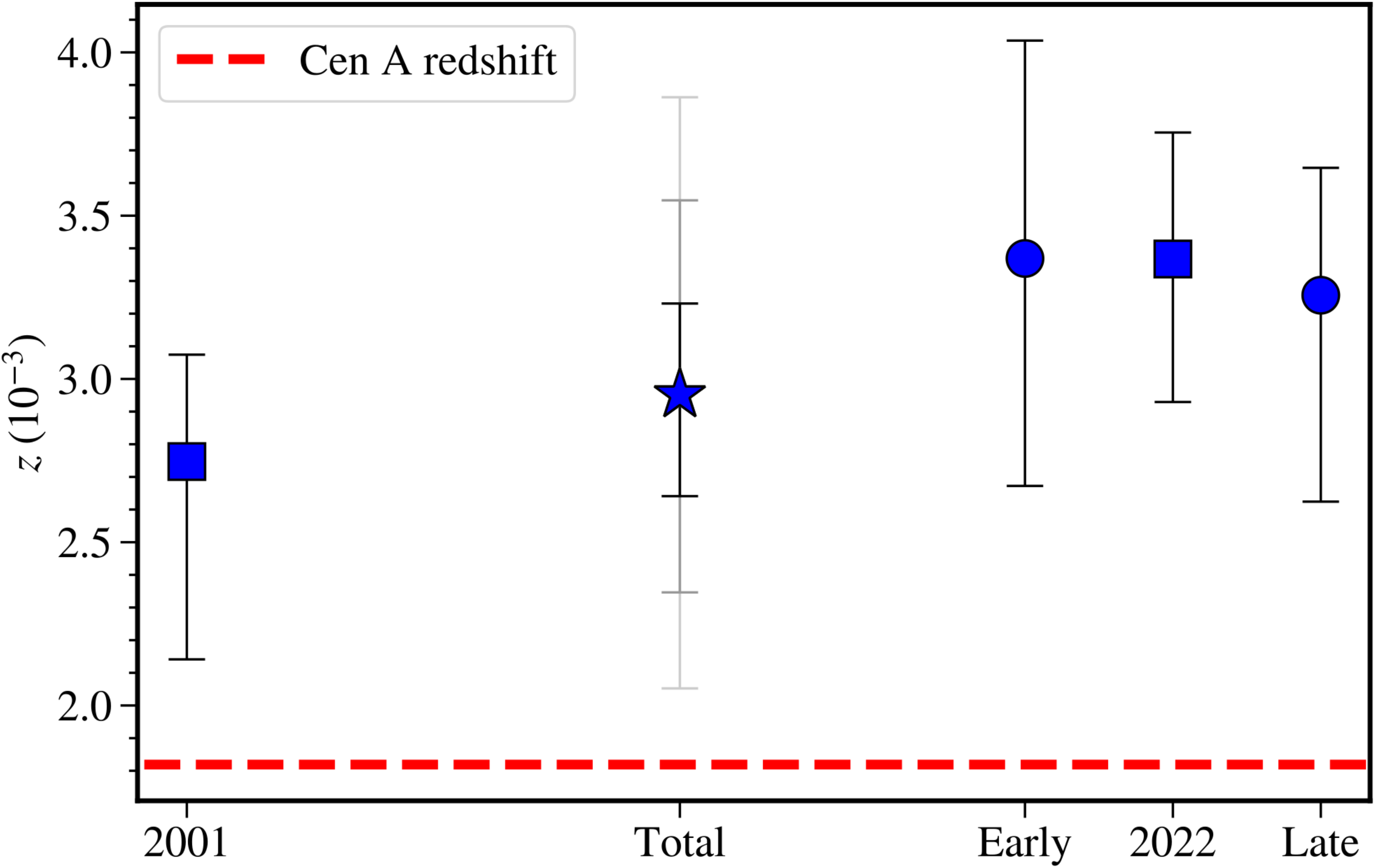}}
\caption{Redshift of the Fe K$\alpha$ line, as found from the best fits using model A to the 2001, early 2022, late 2022, 2022, and total spectra. The red dashed line indicates the redshift of the systemic velocity of Cen A. The redshift for the total spectrum is shown alongside its $1\sigma$ (black), $2\sigma$ (dark grey), and $3\sigma$ (light grey) error bars. 
 \label{fig:z}}
\end{figure}

The Fe K$\alpha$ line is consistently found at a higher redshift that is inconsistent with that of the galaxy as a whole, as is shown in Fig. \ref{fig:z}. The redshifts found in the 2022 spectra are larger than the one from the 2001 spectrum, but they are nevertheless all still consistent with each other, within $1\sigma$ errors. The best fit parameters of the Fe K$\alpha$ line remained mostly consistent between 2001 and 2022. This means that the total spectrum provides the best estimate of its properties. The 1, 2, and $3\sigma$ errors of the Fe K$\alpha$ line energy for the total spectrum correspond to redshifts of: $2.95^{+0.28}_{-0.31}\times10^{-3}$, $\pm0.60\times10^{-3}$, and $\pm0.90\times10^{-3}$, respectively. This is also depicted in Fig. \ref{fig:z}. We found that the spectra could not be well fitted by setting the centroid energy to the value expected from the Cen A redshift. Despite the decrease in fit parameters, the BIC increased by 6.8 for the total spectrum, which further demonstrates the inconsistency between the Cen A redshift, and that of the Fe K$\alpha$ line.

\begin{figure}[htp]
\resizebox{\hsize}{!}{\includegraphics{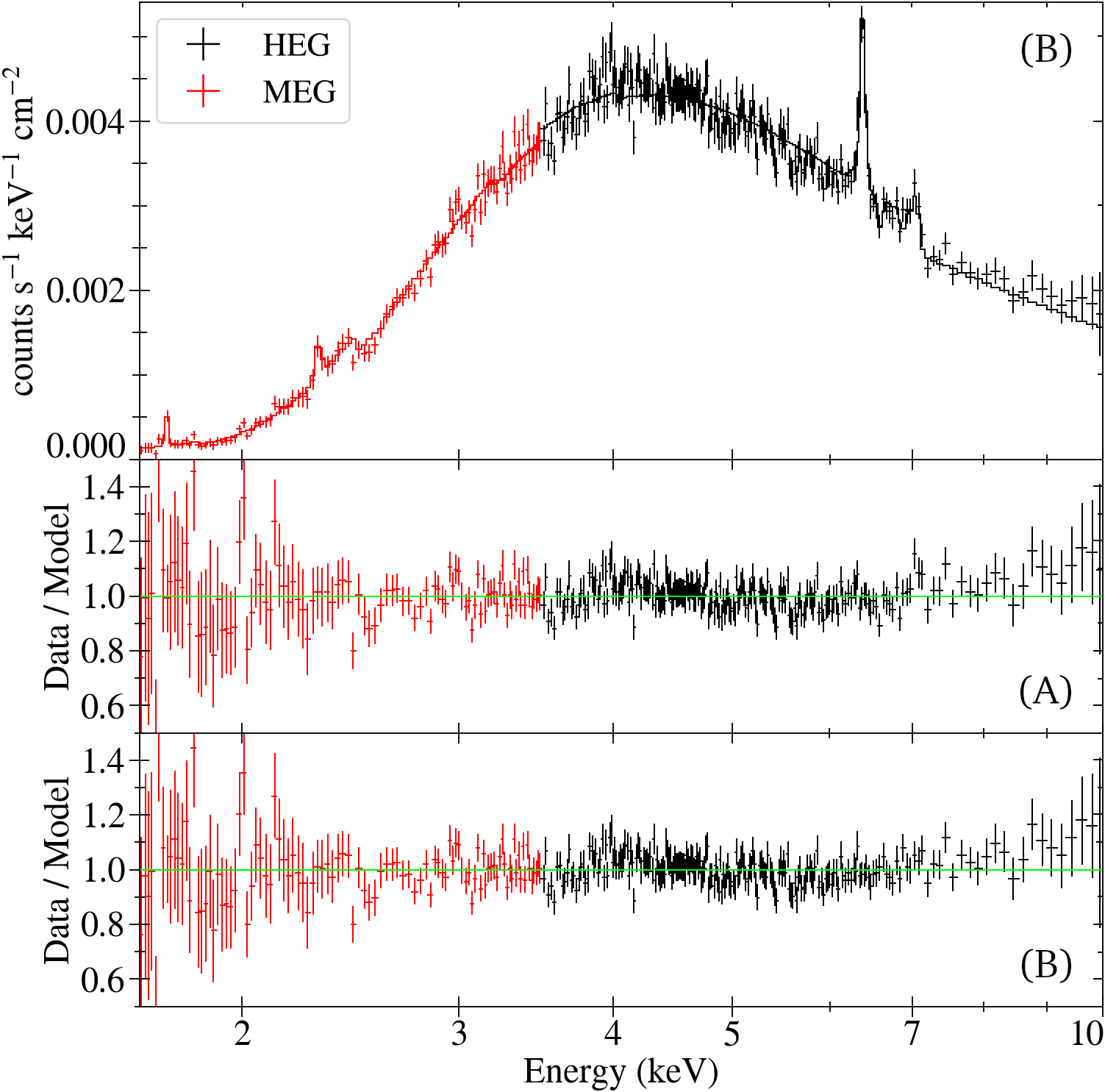}}
\caption{The best fit spectra, and the ratio of the data to the folded model, for the grouped 2022 spectrum. The layout of the spectra is identical to that of Fig. \ref{fig:specmod01}. 
 \label{fig:specmod22}}
\end{figure}

In contrast, the Si K$\alpha$ line in the total spectrum was fitted with a centroid energy that is redshifted by $1.68^{+0.43}_{-0.19}\times10^{-3}$ relative to its rest frame energy. All the measured reshifts of this line are consistent with this value, and also with the Cen A redshift. However, the redshifts of the Si K$\alpha$ line are inconsistent with those found for Fe K$\alpha$. The width of the Si K$\alpha$ line remained constant, within errors, throughout all observations. It increased in amplitude, from a normalization of $33.9\pm4.7\times10^{-4}~\mathrm{photons}~\mathrm{cm}^{-2}~\mathrm{s}^{-1}$ in 2001, to $62.0^{+9.7}_{-9.0}\times10^{-4}~\mathrm{photons}~\mathrm{cm}^{-2}~\mathrm{s}^{-1}$ in late 2022. 

Of the three fluorescent emission lines, the S K$\alpha$ line is the least well constrained in the spectra. Its centroid energy was fit with a wide range of different redshifts, some of which were not consistent with each other. The total spectrum was best fit with a Gaussian function that was redshifted by $z = 1.40^{+0.71}_{-0.92}\times10^{-3}$. This value is consistent with the Cen A redshift, the redshift of the Si K$\alpha$ line, but inconsistent with the redshift of the Fe K$\alpha$ line. There is also significant variation in the best fit width of the S K$\alpha$ line. However, it is unclear if this is caused by a variation in the line, or instead reflects the complex shape of the spectrum at these energies. The amplitude remained consistent within errors.  

We investigated if the Si K$\alpha$ and S K$\alpha$ lines could be self-consistently described by an AGN fluorescent emission line spectrum. Regardless of the ionization degree, it was not possible to get an accurate fit to the spectral shape that included these lines, especially not when also including the Fe K$\alpha$ line. Furthermore, the Si and S lines are too bright compared to the expected ratio of their fluxes relative to Fe \citep{2020ApJ...904...40R}. Rather than the expected $0.104:0.069:1.0$ ratio of the Si to S to Fe fluxes, we found ratios of $5.38\pm0.56:0.506\pm0.095:1.0$ for the total spectrum. Therefore, we conclude that the Si K$\alpha$ and S K$\alpha$ lines in these spectra are not produced by the central engine around the SMBH. This is further supported by the Si and S lines having redshifts that are inconsistent with those measured for Fe K$\alpha$. In contrast, the two lines are consistent with the Cen A redshift, which indicates an accurate energy calibration of the \emph{Chandra} instruments. Therefore, the excess redshift of the Fe K$\alpha$ line is unlikely to have been caused by an error in the energy calibration. 

This could also explain the variation in the unabsorbed Si K$\alpha$ flux, which correlated with the Hydrogen column density. It might have been caused by the wrong assumption that the line is absorbed by the same column density as the rest of the spectrum. When fitting the Si line without the \texttt{ztbabs} absorption component, its flux was found to remain consistent across all spectra. Using a different absorption for the Si and S lines affects their amplitude, but the measured width remains consistent within errors. The lines are shifted to slightly higher energies, which slightly reduces their measured redshifts, but does not affect any of the above discussion of their properties, as they are still consistent with the Cen A redshift. 

The normalization of the Fe K$\alpha$ line was found to vary slightly between 2001 and 2022. This corresponds to a flux decrease of $18.8\pm8.8\%$. In contrast, the continuum flux varied significantly more, by $47.2\pm0.2\%$. As a result of this, the equivalent width of the line is largest when the continuum flux is lowest, so in the early 2022 grouped spectrum. The normalization of the Fe K$\alpha$ line remained stable between early and late 2022, indicating that it varies more slowly, and possibly to a lesser degree than the continuum flux. 

We also fitted spectral model A, but with the \texttt{diskline} component replaced with \texttt{gauss}, a Gaussian line profile. The results of these fits are very similar to those shown in Table \ref{tab:modA1}. In particular, the redshifts for the Fe K$\alpha$ line were identical, within errors, and were also all inconsistent with the Cen A redshift. In these fits, we found the width of the line, as parameterized through the standard deviation of the Gaussian to increase from $\sigma = 18.8^{+4.6}_{-2.8}~\mathrm{eV}$ in 2001, to $\sigma = 28.9^{+2.9}_{-2.3}~\mathrm{eV}$ in 2022. This corresponds to a velocity dispersion for the half width at half maximum of between  $1.04^{+0.25}_{-0.16}\times10^6~\mathrm{m}~\mathrm{s}^{-1}$ and $1.59^{+0.16}_{-0.13}\times10^6~\mathrm{m}~\mathrm{s}^{-1}$. This is broader than the energy resolution of the HETG, and can therefore not be purely attributed to it. The shift, and varying width of the Fe K$\alpha$ line can be seen in Fig. \ref{fig:specFEreg}, which zooms into the $6.15-7.15~\mathrm{keV}$ interval of the spectra. It also shows the asymmetry of the line, which is expected.

The variable width of the Fe K$\alpha$ line is described by a variable emissivity, when fit with \texttt{diskline}. The emissivity power law index decreased from $-1.96\pm0.15$ in 2001, to $-2.40^{+0.14}_{-0.09}$ in 2022. The inner radius of the accretion disk was found to be consistent within errors from 2001 to 2022, possibly due to a degeneracy with the emissivity.  

Motivated by the energy shift of the Fe K$\alpha$ line, we also investigated whether the redshift used by \texttt{ztbabs} differs from the Cen A value. In so doing, we investigated to what extent the absorption edges were shifted in energy. Allowing this redshift to vary freely only slightly reduced the C statistic, but always increased the BIC. Although the best fit redshifts of \texttt{ztbabs} were slightly larger than the Cen A redshift, they had large errors, which made them consistent with it. Therefore, we kept the \texttt{ztbabs} redshift frozen at a value of $1.819\times10^{-3}$ in the best fit results shown in Tables \ref{tab:modA1} and \ref{tab:modB1}, as well as Fig. \ref{fig:parcompAB}. 

\subsection{Model B}

\begin{figure}[h]
\resizebox{\hsize}{!}
{\includegraphics{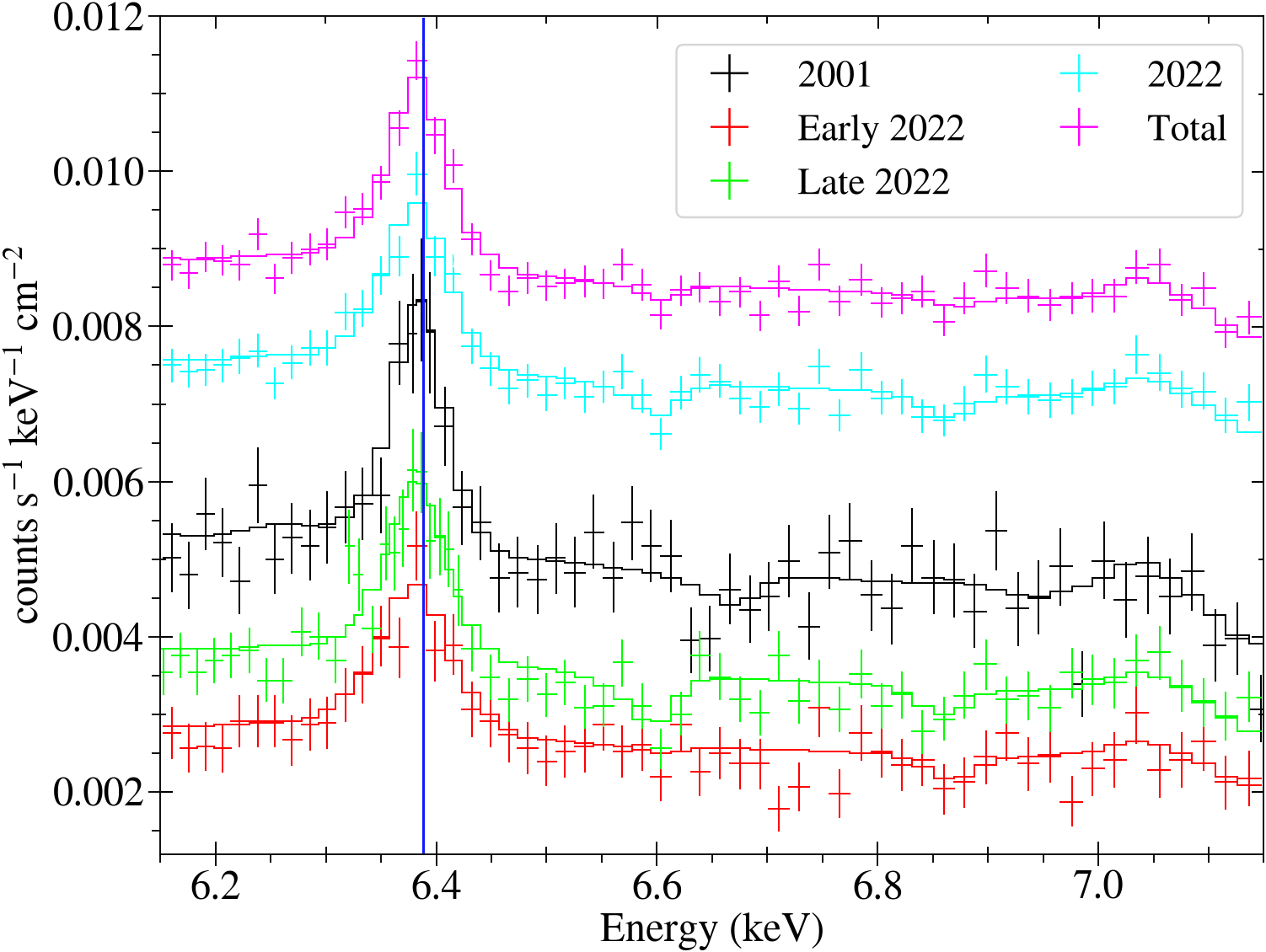}}
\caption{Zoomed in spectra of the $6.15-7.15~\mathrm{keV}$ energy range, featuring the Fe K$\alpha$ and Fe K$\beta$ emission lines, the Fe XXV and Fe XXVI absorption lines, as well as the Fe edge. The best fits of model B to the spectra are also depicted as continuous lines through the data. The vertical blue line describes the expected centroid energy of the Fe K$\alpha$ line, based on the Cen A systemic velocity. The spectra were rebinned for display clarity. The 2022 and total spectra were shifted upwards in this figure, to distinguish them from the other spectra.
 \label{fig:specFEreg}}
\end{figure}

\noindent
In the interval between the Fe K$\alpha$ line and the Fe edge, we detected ionized absorption features corresponding to Fe XXV and Fe XXVI (see Fig. \ref{fig:specFEreg}). To model these absorption lines self-consistently, we used the ``xstar2xspec'' functionality within the XSTAR suite \citep{1982ApJS...50..263K, 2001ApJS..134..139B} to create a table model of photoionized absorption spectra.  The input spectrum was assumed to be typical of an AGN, composed of a $T=25,000~\mathrm{K}$ blackbody and a $\Gamma = 1.7$ power law. The power law component was bent to zero flux below 0.3 keV to avoid infinite flux at low energy. Solar elemental abundances relative to Hydrogen and a gas turbulent velocity of $v_{\rm turb} = 300~{\rm km}~{\rm s}^{-1}$ were also assumed.  The grid includes 6400 individual XSTAR realizations, with 100 grid points in the ionization range $1 \leq {\rm log}~\xi\leq 6$ and 64 grid points in the column density range $1.0\times 10^{21}~{\rm cm}^{-1} \leq  N_{\rm H}\leq 6.0\times 10^{23}~{\rm cm}^{-2}$.  XSPEC is able to interpolate between these models to derive the overall best fit model.  The relatively high resolution of the grid ensures that important gas effects tied to ionization are not lost owing to coarse sampling. In the following, we will refer to this multiplicative model as \texttt{xstar\_abs}.

We also replaced the \texttt{powerlaw} and \texttt{diskline} components of model A with comparable MYTorus \citep{2009MNRAS.397.1549M} components. The zeroth order power law continuum in model B is described by  \texttt{zpowerlw}. The scattered continuum is parameterized by \texttt{constant\textsubscript{s}*mytorusS}, and the fluorescent Fe K$\alpha$ and Fe K$\beta$ lines are described by \texttt{constant\textsubscript{l}*rdblur*mytorusL}. The component \texttt{mytorusL} includes the Compton shoulder, and therefore allows us to test its impact on the measured excess redshift of the Fe K$\alpha$ line. Therefore, model B is described as: \texttt{constant\textsubscript{1}*(tbabs*ztbabs*(xstar\_abs*zpowerlaw+ constant\textsubscript{s}*mytorusS+constant\textsubscript{l}*rdblur*mytorusL+ gauss+gauss)+constant\textsubscript{2}*tbabs*zpowerlaw)}, in XSPEC notation. The parameters of the three MYTorus models were linked, and the photon index, redshift, and normalization were set equal to the values in the \texttt{zpowerlaw} model. As in model A, the two \texttt{gauss} components represent the Si and S K$\alpha$ lines, and \texttt{constant\textsubscript{2}*tbabs*zpowerlaw} describes the leaked power law. Components of this model with identical names had linked parameters, but different subscripts differentiate components with independent parameters. We again assumed an inclination of $60\degree$. 

Table \ref{tab:modB1}, and Fig. \ref{fig:parcompAB} show the best fit results when fitting the five spectra with model B. Identical parameters had mostly the same values, within errors, between models A and B. The best fit parameters of the Si and S K$\alpha$ lines are consistent between the two models, and have not been repeated in Table \ref{tab:modB1}.

In model B, we assumed that the ionized absorber, \texttt{xstar\_abs}, is located closer to the central engine than the origin of the reflection and fluorescence spectrum, and therefore only acts on the coronal spectrum, \texttt{zpowerlaw*mytorusZ}.  To account for velocity differences and emission and absorption at different radii, we allowed the redshift of the ionized absorber to vary.

The properties of the ionized absorber varied significantly from 2001 to 2022. The ionized column density increased alongside the Hydrogen column density, from $0.50^{+0.30}_{-0.24}\times10^{22}~\mathrm{cm}^{-2}$, to $1.38^{+0.38}_{-0.32}\times10^{22}~\mathrm{cm}^{-2}$. At the same time, its redshift increased by a factor of $3.5\pm1.4$, from $4.2\pm1.7\times10^{-3}$, to $15.0^{+0.9}_{-0.7}\times10^{-3}$. This redshift even exceeds that of the Fe K$\alpha$ line, in all spectra.

The Fe XXV absorption line could not be constrained in the early 2022 spectrum (see Fig. \ref{fig:specFEreg}), so the best fit identified a higher ionization degree for it than for the other spectra. Therefore, the ionization degree and ionized column density of the early 2022 spectrum might be overestimated. The ionization degrees of all other spectra are consistent, within errors. Due to significant variation in the properties of the ionized absorber from 2001 to 2022, its best fit parameters in the fit to the total spectrum are unreliable.

The redshifts of the \texttt{zpowerlaw} component, which are also used by all MYTorus components, are lower than the corresponding redshifts found for the Fe K$\alpha$ line with \texttt{diskline} in model A. This may be the consequence of the use of this redshift to describe many spectral features, not just the Fe K$\alpha$ line. Therefore, the redshifts found in model A are still the most reliable for estimating the shift of the Fe K$\alpha$ line itself. In the fits with model B, the redshifts for the 2001 and early 2022 spectra are comparatively low and high, and have strongly asymmetric uncertainties. Nevertheless, the redshifts found by model B are still consistent with the values found by model A, and are inconsistent with the Cen A redshifts, for all five spectra. This further supports the notion that the Fe K$\alpha$ line is indeed found at a lower energy than expected from the Cen A redshift, even when including a Compton shoulder. The components of the \texttt{rdblur} model are consistent, yet less well constrained than the comparable parameters of the \texttt{diskline} model from spectral model B. 

The MYTorus $N_{\rm H}$ describes the column density in the equatorial plane of the torus. At an inclination of $60\degree$, it does not contribute to the absorption seen along the line of sight, but instead affects the fluorescent and reflection features. That is why a \texttt{mytorusZ} component was not included, as it has no effect at this inclination. The MYTorus Hydrogen column density appeared to vary significantly from 2001 to 2022. However, the reason for this is that it could not be well constrained for the 2001 spectra, and was likely to have been significantly overestimated, as it only has a small impact on the spectra at an inclination of $60\degree$. 

Of particular note is that the strength of the scattering component, parameterized by the \texttt{constant\_s} factor, was found to be 0 for all spectra. We could only place an upper limit on this reflection fraction, of $c_s < 0.027$ for the total spectrum. The variation of the \texttt{constant\_l} factor is the result of changes in other parameters, and does not reflect the small variation in the Fe K$\alpha$ line flux (see Table \ref{tab:modA1}).

The spectral fits using model B have a lower C statistic than the ones using model A, but have a higher BIC. Model A is the simplest possible model to describe the main spectral shape. Model B is more complex, and uses more parameters to describe finer features of the spectra. 

\begin{table*}[ht]
\centering
\setlength{\tabcolsep}{3pt}
\def\arraystretch{1.1}
\begin{tabular}{l l l || l | l l l | l}
    \textbf{Component} & & \textbf{Units} & \textbf{2001} & \textbf{Early 2022} & \textbf{Late 2022} & \textbf{2022} & \textbf{Total} \\ \hline \hline
    \texttt{ztbabs} & $\boldsymbol{N_{\rm H1}}$ & $10^{22}~\mathrm{cm}^{-2}$ & $15.30\pm0.05$ & $16.53^{+0.08}_{-0.07}$ & $16.94\pm0.07$ & $16.78\pm0.05$ & $16.05\pm0.04$ \\ \hline
    \texttt{xstar\_abs} & $\boldsymbol{N_{\rm I}}$ & $10^{22}~\mathrm{cm}^{-2}$ & $0.50^{+0.30}_{-0.24}$ & $2.7^{+1.7}_{-1.5}$ & $1.57^{+0.35}_{-0.48}$ & $1.38^{+0.38}_{-0.32}$ & $0.56^{+0.28}_{-0.25}$ \\ 
    & $\boldsymbol{\log(\xi)}$ & & $4.22^{+0.11}_{-0.21}$ & $4.84^{+0.37}_{-0.19}$ & $4.31^{+0.08}_{-0.07}$ & $4.38^{+0.09}_{-0.07}$ & $4.39^{+0.17}_{-0.10}$ \\ 
    & $\boldsymbol{z_{\mathrm I}}$ & $10^{-3}$ & $4.2\pm1.7$ & $14.6^{+1.5}_{-1.3}$ & $15.6^{+0.5}_{-1.1}$ & $15.0^{+0.9}_{-0.7}$ & $14.4^{+1.4}_{-1.1}$ \\ \hline
    \texttt{zpowerlaw} & $\boldsymbol{\Gamma}$ & & $1.636^{+0.003}_{-0.002}$ & $1.811\pm0.003$ & $1.804\pm0.003$ & $1.805^{+0.003}_{-0.002}$ & $1.743^{+0.002}_{-0.001}$ \\ 
    & $\boldsymbol{z}$ & $10^{-3}$ & $2.04^{+0.55}_{-0.04}$ & $3.29^{+0.05}_{-0.94}$ & $2.60^{+0.69}_{-0.57}$ & $2.82^{+0.48}_{-0.26}$ & $2.59^{+0.68}_{-0.54}$  \\ 
    & $\boldsymbol{N_{\rm PL}}$ & $\mathrm{photons}~\mathrm{keV}^{-1}~\mathrm{cm}^{-2}~ \mathrm{s}^{-1}$ & $0.1232^{+0.0004}_{-0.0005}$ & $0.0916^{+0.0035}_{-0.0047}$ & $0.1245^{+0.0005}_{-0.0006}$ & $0.1113^{+0.0003}_{-0.0005}$ & $0.1125^{+0.0003}_{-0.0002}$ \\ \hline
    \texttt{mytorus} & $\boldsymbol{N_{\rm H2}}$ & $10^{22}~\mathrm{cm}^{-2}$ & $240^{+180}_{-70}$ & $32^{+28}_{-6}$ & $41^{+26}_{-9}$ & $40.8^{+8.4}_{-6.8}$ & $33.6^{+5.0}_{-3.7}$ \\ \hline
    \texttt{constant\textsubscript{s}} & $\boldsymbol{c\textsubscript{s}}$ & & $0.00^{+0.088}_{-0.000}$ & $0.00^{+0.10}_{-0.00}$ & $0.00^{+0.05}_{-0.00}$ & $0.00^{+0.035}_{-0.000}$ & $0.000^{+0.027}_{-0.000}$ \\ \hline
    \texttt{constant\textsubscript{l}} & $\boldsymbol{c\textsubscript{l}}$ & & $0.726^{+0.061}_{-0.068}$ & $1.12\pm0.11$ & $0.747^{+0.075}_{-0.070}$ & $0.839\pm0.055$ & $0.817\pm0.045$ \\ \hline
    \texttt{rdblur} & $\boldsymbol{q}$ & & $-1.98^{+0.38}_{-0.15}$ & $-2.25\pm0.16$ & $-2.13^{+0.18}_{-0.19}$ & $-2.18^{+0.13}_{-0.12}$ & $-2.04^{+0.09}_{-0.10}$ \\ 
    & $\boldsymbol{R_{\rm in}}$ & $10^3~r_{\rm g}$ & $8^{+13}_{-3}$ & $4.4^{+2.0}_{-1.4}$ & $6.3^{+3.7}_{-2.2}$ & $5.6^{+1.9}_{-1.3}$ & $4.6^{+2.1}_{-1.0}$ \\  \hline 
    \texttt{constant\textsubscript{2}} & $\boldsymbol{c_2}$ & $10^{-3}$ & $2.77^{+0.20}_{-0.17}$ & $3.05\pm0.27$ & $2.54\pm0.17$ & $2.73\pm0.14$ & $2.93\pm0.12$ \\ \hline \hline
    & $\boldsymbol{C}$ & & $1775.28$ & $1810.77$ & $1786.09$ & $1863.14$ & $1904.46$ \\ 
    & $\boldsymbol{BIC}$ & & $1924.20$ & $1959.69$ & $1935.01$ & $2012.06$ & $2053.38$ \\ 
\end{tabular}
\caption{Best fit properties of fitting the spectra with XSPEC model B, \texttt{constant\textsubscript{1}*(tbabs*ztbabs* (xstar\_abs*zpowerlaw+constant\textsubscript{s}*mytorusS+constant\textsubscript{l}*rdblur*mytorusL+gauss+gauss)+constant\textsubscript{2}*tbabs*zpowerlaw)}. The ionized absorber model, \texttt{xstar\_abs}, has parameters of $N_{\rm I}$, the ionized column density, and $\xi$, the ionization degree of the accretion disk. The \texttt{rdblur} model has an emissivity power law index of $q$, and an inner accretion disk radius of $R_{\rm in}$. All the MYTorus models use the same Hydrogen column density, $N_{\rm H2}$, as well as the parameters of the \texttt{zpowerlaw} component. The remaining parameters are described in Table \ref{tab:modA1}. The parameters of the two \texttt{gauss} components are equivalent to those listed in Table \ref{tab:modA1}. For all of these fits, there are 1713 bins, and 1694 degrees of freedom.}  
\label{tab:modB1}
\end{table*}

\section{Discussion} \label{sec:dscsn}

\subsection{Neutral absorption, power law}

\noindent
Comparing the different spectral fits, we investigated how variable individual components were over the course of months and decades. The main spectral shape remained mostly consistent throughout all observations, and the largest variation was observed in the amplitude of the power law. 

The Hydrogen column density was observed to vary from early to late 2022, and featured even larger differences when compared over a 21-year interval. This result is expected, and consistent with previous findings. 

The photon index of the power law was found to be consistent at a value of $\approx1.81$ throughout 2022. This is also consistent with many previous spectral analyses that measured a similar photon index \citep{1978QJRAS..19....1C, 2003ApJ...593..160G, 2006ApJ...641..801R, 2011ApJ...733...23R, 2016ApJ...819..150F}. However, the merged 2001 spectrum was found to be significantly shallower, with $\Gamma=1.646\pm0.002$, using model A. This result is in agreement with that of \citet{2004ApJ...612..786E}, and corresponds to a similarly shallow slope as the one found by \citet{1978ApJ...220..790M} and \citet{1981ApJ...244..429B}. The inability to fit all \emph{Chandra} spectra from 2001 and 2022 with the same photon index provides an indication that it is variable, and can become significantly shallower than it usually is, albeit on timescales of years or decades. 

Furthermore, we detected an anticorrelation between the photon index and the luminosity of Cen A. The 2001 spectrum was both the hardest, with a photon index of $\Gamma=1.646\pm0.002$ (for model A), and the brightest, with an unabsorbed $2-10~\mathrm{keV}$ luminosity of $1.40\pm0.26\times10^{-4}~L_{\rm Edd}$. In contrast, the 2022 spectrum was best fit with $\Gamma = 1.803\pm0.002$, and an unabsorbed luminosity of $9.7\pm1.8\times10^{-5}~L_{\rm Edd}$. This anticorrelation agrees with the results of \citet{2015MNRAS.447.1692Y} and \citet{2016MNRAS.459.3963C}, who found a decreasing photon index with an increasing luminosity for AGN accreting with a $2-10~\mathrm{keV}$ luminosity between $10^{-6.5}-10^{-3}~L_{\rm Edd}$. A possible explanation for this effect is that the synchrotron power law from the jet becomes stronger than that from the advection-dominated accretion flow \citep{2005ApJ...629..408Y}.

\subsection{Fe K$\alpha$ line}
\noindent
Several previous studies of the Cen A X-ray spectrum concluded that the disk generating the fluorescent Fe K$\alpha$ line had to have a large extent to account for the apparent stability of its flux over intervals of several years, compared with a significant variability of the continuum \citep{2006ApJ...641..801R, 2011ApJ...733...23R, 2016ApJ...819..150F}. We found the Fe K$\alpha$ flux to only vary slightly, by $18.8\pm8.8\%$ between 2001 and 2022. This is comparable to the $20-30\%$ variation found by \citet{2011ApJ...743..124F}. However, it should be noted that the flux we found for the line in the total spectrum, $2.08^{+0.14}_{-0.12}\times10^{-4}~\mathrm{photons}~\mathrm{cm}^{-2}~\mathrm{s}^{-1}$, is less than half as large as the consistent $4.55\pm0.14\times10^{-4}~\mathrm{photons}~\mathrm{cm}^{-2}~\mathrm{s}^{-1}$ flux found by \citet{2011ApJ...733...23R} using RXTE data from 1996 to 2009. 

The lack of large variation in the Fe K$\alpha$ line flux is contrasted by \citet{2022A&A...664A..46A}, who observed a variation of about an order of magnitude between two non-grating \emph{Chandra} ACIS observations of Cen A. However, we re-analysed the spectra from those two observations, and measured a consistent Fe K$\alpha$ line flux. 

We also found the width of the Fe K$\alpha$ line to vary from 2001 to 2022, but remain consistent on short time scales. The increased standard deviation of the best fit \texttt{gauss} model corresponds to an increased emissivity and a slightly larger, albeit still consistent inner radius, when fitting with \texttt{diskline}.

The centroid energy of the Fe K$\alpha$ line was observed to be significantly offset from its expected energy based on the known recession velocity of Cen A. In Model A, we measured a redshift of $2.95^{+0.28}_{-0.31}\times10^{-3}$ for the total spectrum, compared to the Cen A redshift of $1.819\pm0.010\times10^{-3}$. This excess redshift was observed by both spectral models A and B, as well as other models we investigated, involving \texttt{gauss}, and \texttt{pexmon} components. The excess redshift also does not depend on the assumption made in the spectral fits, such as the selection of an inclination of $60\degree$. The Fe K$\alpha$ line energy was found to only vary within the respective errors from 2001 to 2022.

AGN spectral lines between $6$ and $7~\mathrm{keV}$ with significant redshifts relative to the systemic velocity of the galaxy, have been previously found by \emph{Chandra} in $\mathrm{M81}^{\ast}$ \citep{2007ApJ...669..830Y, 2021NatAs...5..928S}. However, these concerned the Fe XXVI Ly$\alpha$ emission line, and also featued a blueshifted line.  

The absolute energy calibration of \emph{Chandra}'s HETGS has a systematic error on the scale of $\approx 100 ~ \mathrm{km} ~ \mathrm{s}^{-1}$ \footnote{\href{https://cxc.harvard.edu/proposer/POG/html/chap8.html}{https://cxc.harvard.edu/proposer/POG/html/chap8.html}}. Doppler shifts caused by velocities of as low as $\approx 50 ~ \mathrm{km} ~ \mathrm{s}^{-1}$ have been detected \citep{2012MNRAS.421.3550Z}. Furthermore, both the Si and S K$\alpha$ lines, which probably do not originate in the accretion disk, were redshifted by an amount slightly smaller, but still consistent with the Cen A redshift, and inconsistent with the Fe K$\alpha$ redshift. This leads us to conclude that the excess redshift of the Fe K$\alpha$ line is likely not the result of an offset of the \emph{Chandra} energy calibration.  

The redshift to Cen A is measured relative to a heliocentric reference frame. The motion of \emph{Chandra} relative to this reference frame might slightly offset the measured energies. We assume that the Doppler shift caused by the \emph{Chandra} orbit around Earth, averaged over several observations, is small compared with the shift caused by the orbit of Earth around the Sun. The magnitude of the component of the orbital velocity of Earth in the direction of Cen A, as observed in the heliocentric reference frame, is at most $25.4~\mathrm{km}~\mathrm{s}^{-1}$. The average orbital velocity component in the direction of Cen A, weighted by the exposure time, was calculated to be $25.3$, $15.8$, $4.5$, $7.9$, and $13.2~\mathrm{km}~\mathrm{s}^{-1}$, for the 2001, early, and late 2022, 2022, and total observations, respectively. Subtracting these from the measured redshifts, we find the Fe K$
\alpha$ line to have an average radial velocity relative to the Cen A systemic velocity, of $250^{+100}_{-180}$, $450\pm200$, $430^{+120}_{-190}$, $460^{+120}_{-130}$, and $326^{+84}_{-94}~\mathrm{km}~\mathrm{s}^{-1}$, respectively. Using the XSPEC \texttt{error} command, we found the line to be offset from the Cen A redshift by $3.62\sigma$, equivalent to a p-value of $0.0145\%$, for the total spectrum, fit with model A. We found a similar significance with model B as well.

These velocity shifts are still up to one order of magnitude smaller than the component of the orbital velocity in the line of sight at the inner radius. The width of the line is far greater than the shift of its center away from the Cen A systemic velocity. 

Studies of the varying kinematics in the images of infrared and radio spectral lines of the circumnuclear disk within several hundred pc of the SMBH revealed complex structures that have been explained via a warped disk model \citep{2006ApJ...645.1092Q, 2007ApJ...671.1329N, 2017ApJ...843..136E, 2017ApJ...851...76M}. These studies traced red-, and blueshifted regions, but did not find any clear indication of large bulk motion relative to the Cen A systemic velocity. However, \citet{2017ApJ...851...76M} found two absorption complexes, one of which moved at the systemic velocity of Cen A, the other was redshifted by $60~\mathrm{km}~\mathrm{s}^{-1}$.

One way to interpret the excess redshift of the Fe K$\alpha$ line, is if the warped structure is still present at much smaller radii. For simulations of warped accretion disks around black holes, see e.g. \citet{1999MNRAS.304..557O, 2014MNRAS.441.1408T, 2023NewA..10102012L, 2023ApJ...944L..48L} In that case, the excess redshift could be caused by a greater visibility of the redshifted side of the disk. The blueshifted side would occupy a smaller solid angle, and parts of it may be blocked by the warp. \citet{2021ApJ...906...28A} simulated the Fe K$\alpha$ line profile for a warped disk, and found that it can appear to be shifted to lower energies. 

In this model, we would expect the warp to propagate around the disk, which would change the size of the excess redshift identified. At a radius of $5\times10^3~r_{\rm g}$, the orbital period is 19 years. Most of the Fe line originates at greater distances from the SMBH, and the warp is expected to have a longer precession than orbital period \citep{2012PASJ...64...40I}. Therefore, this model can account for the consistency of the excess redshift over 21 years, but does require it to vary sinusoidally on longer timescales.

A theoretical study by \citet{1996MNRAS.281..357P} argued that AGN disks would only show warps at radii of $\geq 10^6~r_{\rm g}$. However, in a followup study, the critical radius above which disks can develop warps was set at $r_{\rm crit} \approx 4\times10^3 ~ r_{\rm g}$ \citep{1997MNRAS.292..136P}. Nevertheless, large amplitude warps were only found to develop at radii orders of magnitude larger than the critical radius. The characteristic timescale of variability of the warp was calculated to be of order $10^6~\mathrm{yr}$. Therefore, it remains unclear if significant warps can develop at the required radii to account for the observed excess redshift.  

A different class of explanations for the observed excess redshift can be found by arguing that it reflects the bulk motion of the central engine relative to the center of the galaxy. This motion might be in the form of an oscillation with a period that is many times longer than the span of the observations. Assuming a constant velocity at the value found from the total spectrum, the central engine would have travelled $7.0^{+1.8}_{-2.0}\times 10^{-3}~\mathrm{pc}$ along the line of sight over the 21-year span of observations.

This would be an unusually large velocity of the SMBH relative to its host galaxy. However, even larger velocities have been inferred in a number of systems with off-center SMBHs \citep{2014ApJ...796L..13M, 2016ApJ...817..150M, 2019ApJ...885L...4S, 2020ApJ...888...36R, 2023MNRAS.522..948C}. Velocities like these can be caused from the merger of two SMBHs. Asymmetries in the spin and mass of merging SMBHs result in an asymmetric gravitational wave emission, which produces a recoil of the merged SMBH \citep{1962PhRv..128.2471P, 2008ApJ...678..780G, 2011MNRAS.412.2154B, 2016MNRAS.456..961B}, with velocities of up to $4000~\mathrm{km}~\mathrm{s}^{-1}$ \citep{2007ApJ...659L...5C}. This recoil results in a damped oscillation around the galactic center that can last for more than $1 ~\mathrm{Gyr}$ \citep{2008ApJ...678..780G}. The kinematics, metallicities, and halo features of Cen A suggest that it had a major merger $\approx 2~\mathrm{Gyr}$ ago \citep{2020MNRAS.498.2766W}. For a large velocity of the SMBH relative to the galaxy to be maintained for $2~\mathrm{Gyr}$ would require a large initial velocity, and weak dynamical friction.

The main drawback of this interpretation is that the AGN spectral lines observed at other wavelengths do not feature a comparably large offset from the Cen A systemic velocity. Additionally, it is unusual for an SMBH to have a large velocity relative to the galaxy, but still be found at its center. 

One way of maintaining a large velocity at small radii, but only minimal bulk motion at distances of hundreds of parsecs, is if the velocity shift is caused by the orbit of the SMBH around another massive body. Given the consistency of the measured excess redshift, this would have to be a large orbit, with an orbital period much longer than 21 years. Given the size of the velocity shift, the secondary body would also have to be comparably massive, so it would have to be another SMBH. A close binary SMBH in Cen A was also suggested by \citet{2022Ap&SS.367...92C} to explain peculiarities in the \emph{EHT} image of the galactic center of Cen A \citep{2021NatAs...5.1017J}. However, \citet{2022Ap&SS.367...92C} argued for orbital periods in the $10^{-1} - 10^1~\mathrm{yrs}$ order of magnitude range, which might be too small to justify the consistency of the excess redshift observed. 

A third type of explanation for this shift, is that it reflects an outflow of material from the disk. Cen A hosts prominent jets that present one avenue for an outflow, albeit at a higher velocity \citep{1998AJ....115..960T, 2019ApJ...871..248S}. If the excess redshift of the Fe K$\alpha$ line is caused by an outflow, it would require the line emission to predominantly originate from the far side of the black hole. The emission from the near side would either have to be suppressed, or be from non-outflowing material. In this scenario, we might expect the excess redshift to vary significantly over the course of a few years. It remains unclear whether such a model could account for the consistency observed over 21 years. Asymmetric AGN outflows resulting in a redshifted Fe XXVI emission line have previously been described by \citet{2007ApJ...669..830Y}. However, \citet{2021NatAs...5..928S} later observed both redshifted and blueshifted components of the line for the same source. Similarly, the excess redshift could alternatively represent a inflow of material towards the SMBH on the closer side. It is similarly unclear if such an inflow could remain consistent over 21 years. 

Another possibility is that the Fe K$\alpha$ line is produced by gas illuminated by the counterjet, and accelerated by it to the measured recession velocity. This could account for the excess redshift, the equivalent width of the line, and the consistency in time. However, it has difficulty explaining the velocity dispersion observed in the line, which is up to an order of magnitude larger than the shift. This model would also require that a similar region does not exist in the direction of the jet, or that the emission from it is strongly absorbed. This could be possible if it has a high column density, thereby obscuring the emission.

Further studies of the excess redshift of the Fe K$\alpha$ line, and its variability over years and decades are required to distinguish between the different potential explanations that were discussed here. A greater spectral resolution could also help to identify possibly suppressed blueshifted wings of the line, and line profiles inconsistent with a planar disk. Investigating the cause of the excess redshift could help improve our understanding of the kinematics in accretion disks around AGNs, and expand the ways in which it can be probed. 

The size of the disk emitting the Fe K$\alpha$ line photons can be estimated from the spectral fits of the line, under the assumption of a particular inclination. When fixing it at a value of $60\degree$, and setting the outer radius to be $10^6 ~r_{\rm g}$, we find consistent inner radii for all grouped spectra. The total spectrum was best fit with $R_{\rm in} = 4.8^{+0.9}_{-1.2}\times10^{3}~r_{\rm g}$. Alternatively, when equating the half width at half maximum of the Fe K$\alpha$ line with the Doppler shift of a stable orbit, for an inclination of $60\degree$, we find radii of $4.1^{+1.0}_{-1.6}\times10^4~r_{\rm g}$, and $2.66^{+0.43}_{-0.54}\times10^4~r_{\rm g}$ for the 2001, and 2022 grouped spectra, respectively. 

\subsection{Fe XXV and Fe XXVI absorption lines}

\noindent
The Fe XXV and Fe XXVI absorption lines at $6.697~\mathrm{keV}$ and $6.966~\mathrm{keV}$ changed significantly from 2001 to 2022. Not only did the ionized absorber column density increase from $0.58\pm0.28\times10^{22}~\mathrm{cm}^{-2}$ to  $1.62\pm0.37\times10^{22}~\mathrm{cm}^{-2}$, the redshift also increased from $4.2\pm1.6\times10^{-3}$ to $15.0^{+0.9}_{-0.7}\times10^{-3}$. In contrast, the ionization degree remained consistent over the 21-year interval.

These redshifts are significantly larger than those found for the Fe K$\alpha$ line. This could be indicative of an inflow of the ionized material. Assuming that the SMBH moves with the systemic velocity of Cen A, rather than with the velocity indicated by the excess redshift of the Fe K$\alpha$ line, the inflow velocities equate to: $690\pm460~\mathrm{km} ~ \mathrm{s}^{-1}$ and $3950^{+260}_{-220}~\mathrm{km} ~ \mathrm{s}^{-1}$, for the 2001, and the 2022 spectra, respectively. If this redshift is caused by an inflow, the properties reflect the variable inflow.

The mass accretion rate can be estimated from the properties of the ionized absorber. We extrapolated from the Fe ion column density to describe the total composition of the inflowing material. Furthermore, we assumed an isotropic accretion, and assumed that the inflow starts at the inner radius of the disk. However, the mass accretion rate estimated in this way is six orders of magnitude too large for the measured luminosities. There is further inconsistency with this association due to the system being brighter at a time when a lower ionized absorber column density and inflow velocity was measured. This indicates that this description might be too simplistic. The redshift of the absorption lines may derive a component from the orbital motion, if the blueshifted components are blocked from view. 

The spectral fits did not require the addition of a broad Fe line at $6.8~\mathrm{keV}$, as suggested by \citet{2003ApJ...593..160G}. There was no indication of a break in the power law. The best fit reflection strength, not including any fluorescent features, was 0 in all spectra, which agrees with the results of \citet{2007ApJ...665..209M, 2011ApJ...733...23R, 2016ApJ...819..150F}, but contrasts those of \citet{2011ApJ...743..124F}. It is unclear why these spectra lack reflection features, as the optical depth should be sufficient to produce these. 

\subsection{XRISM}

\noindent
The Resolve instrument on the X-Ray Imaging and Spectroscopy Mission \citep[\emph{XRISM};][]{2020arXiv200304962X} will be able to determine the properties of the Cen A spectrum with far greater sensitivity than is possible with previous X-ray spectrometers. In particular, it will enable an investigation into the exact properties of the Compton shoulder of the Fe K$\alpha$ line. We simulated a $100~\mathrm{ks}$ \emph{XRISM} spectrum based on the best fit to the total \emph{Chandra} spectrum, using model B. Due to its higher energy resolution, and greater effective area, it will be able to constrain the redshift of the Fe K$\alpha$ line, and the ionized absorber up to one order of magnitude better than was possible when combining $378~\mathrm{ks}$ of \emph{Chandra} HETGS exposure. Furthermore, it will be possible to constrain the inclination of the disk to within $10\degree$, by fitting the exact shape of the Fe K$\alpha$ line at a greater energy resolution. These spectra will also be much more sensitive to the Hydrogen column density in the torus, and the inner radius of the disk.

\section{Conclusions} \label{sec:conc}

\noindent
The AGN at the center of Cen A was observed by the \emph{Chandra} HETGS twice in 2001, and 14 times in 2022. All spectra were well described by an absorbed power law with a strong and narrow Fe K$\alpha$ line. We grouped spectra from different observations together, to analyse weaker spectral features, and investigate spectral variability on timescales of months and years. Variation in the flux on short and long timescales was predominantly caused by a variation in the amplitude of the entire spectrum. 

Within these observations, the AGN was brightest in 2001, and also had the hardest power law slope and the smallest Hydrogen column density. The Hydrogen column density was found to vary on timescales of months. The power law slope varied as well from 2001 to 2022, but remained consistent throughout 2022. 

There was no indication of any reflection features, or any break in the power law. To fit the part of the spectra below $2~\mathrm{keV}$ required the addition of a leaked power law component, with an amplitude of about $0.3\%$ of the main power law feature. 

The Fe K$\alpha$ line increased in width from 2001 to early 2022, and became dimmer by $18.8\pm8.8\%$, but remained consistent throughout 2022. Si and S K$\alpha$ lines were also detected, but could not be associated with the accretion disk.

The energy of the Fe K$\alpha$ line was measured to be lower than expected from the Cen A redshift. This excess redshift was consistently found in different spectral models of the line profile, and in all spectra. The total spectrum was best fit with an excess velocity of $326^{+84}_{-94}~\mathrm{km}~\mathrm{s}^{-1}$, and is inconsistent with the Cen A redshift with a significance of $3.62\sigma$. 

We interpret this result as possibly indicative of a warped accretion disk on sub-parsec scales, which enhances the redshifted, but reduces or obscures part of the blueshifted wing of the line emitting region. We also consider the possibility of motion of the SMBH relative to the center of the galaxy, as well as an outflow or inflow from the disk. 

The spectra also featured absorption lines of Fe XXV and Fe XXVI. The properties of these lines varied significantly from 2001 to 2022, with a higher ionized column density, and a significantly higher redshift in the latest observations. These result may be interpreted as a variable inflow with velocities of $690\pm460~\mathrm{km} ~ \mathrm{s}^{-1}$ and $3950^{+260}_{-220}~\mathrm{km} ~ \mathrm{s}^{-1}$ in 2001 and 2022. 

\section{Acknowledgements}

\noindent
This research has made use of data obtained from the \emph{Chandra} Data Archive and the \emph{Chandra} Source Catalog, and software provided by the \emph{Chandra} X-ray Center (CXC) in the application packages CIAO and Sherpa. This work made use of the software packages astropy \citep[\href{https://www.astropy.org/}{https://www.astropy.org/};][]{2013A&A...558A..33A, 2018AJ....156..123A, 2022ApJ...935..167A}, numpy \citep[\href{https://www.numpy.org/}{https://www.numpy.org/};][]{harris2020array}, matplotlib \citep[\href{https://matplotlib.org/}{https://matplotlib.org/};][]{Hunter:2007}, and scipy \citep[\href{https://scipy.org/}{https://scipy.org/};][]{2020SciPy-NMeth}. EB is partially supported by a Center of Excellence of THE ISRAEL SCIENCE FOUNDATION (grant No. 2752/19). EK acknowledges financial support from the Centre National d’Etudes Spatiales (CNES). SRON is supported financially by NWO, The Netherlands Organization for Scientific Research. AZ is supported by NASA under award number 80GSFC21M0002.

We thank the anonymous referee for their insightful comments.

\bibliography{bibliography}{}
\bibliographystyle{aasjournal}

\appendix

\section{Further figures of spectra fitted with models A and B}\label{sec:app1}

\begin{figure}[htp]
\resizebox{0.96\hsize}{!}{\includegraphics{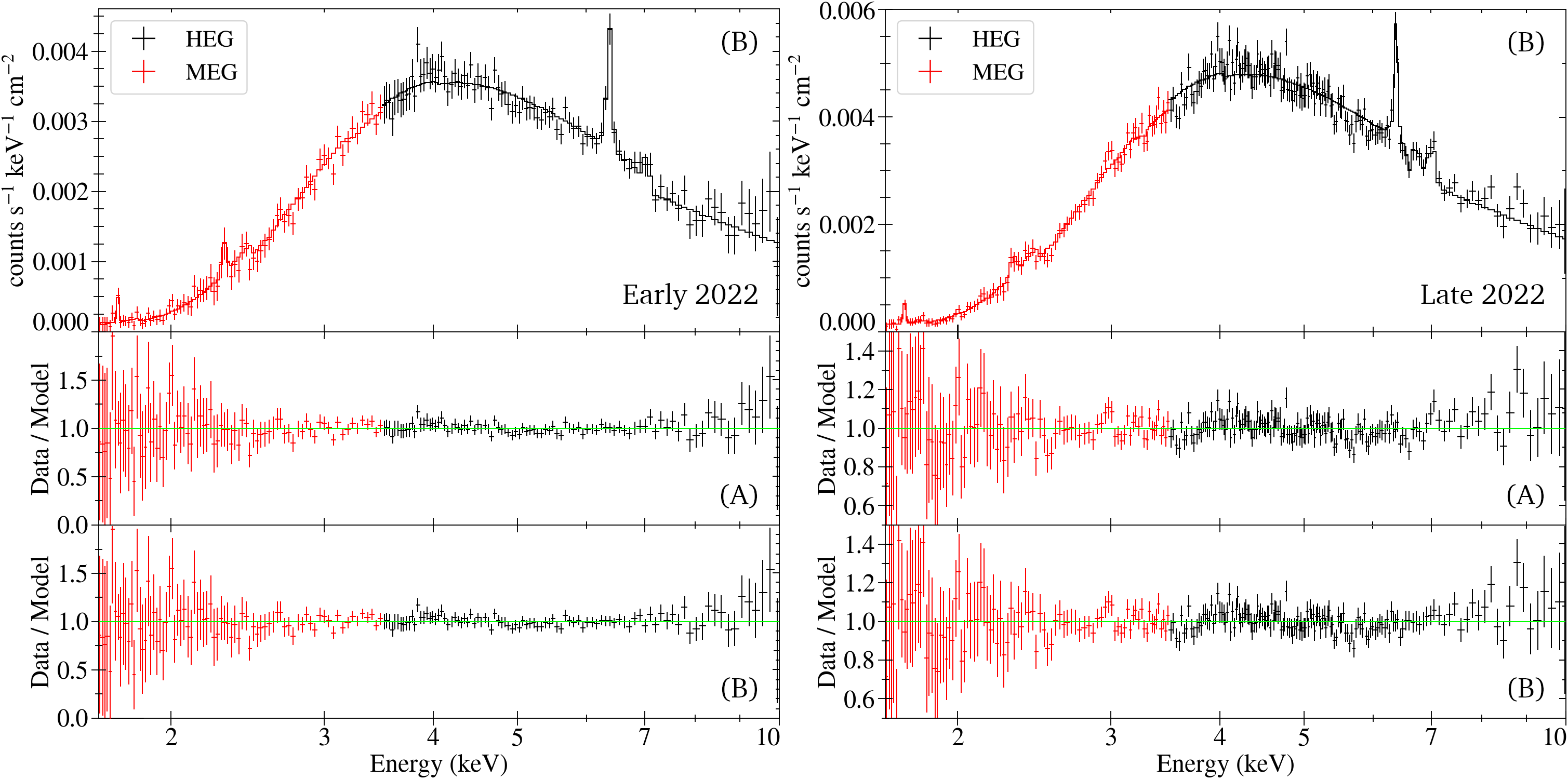}}
\caption{The best fit spectra, and the ratio of the data to the folded model, for the grouped early (left), and late 2022 (right) spectra. The layout of the spectra is identical to that of Fig. \ref{fig:specmod01}. 
 \label{fig:specmod22el}}
\end{figure}

\begin{figure}[htp]
\resizebox{0.48\hsize}{!}{\includegraphics{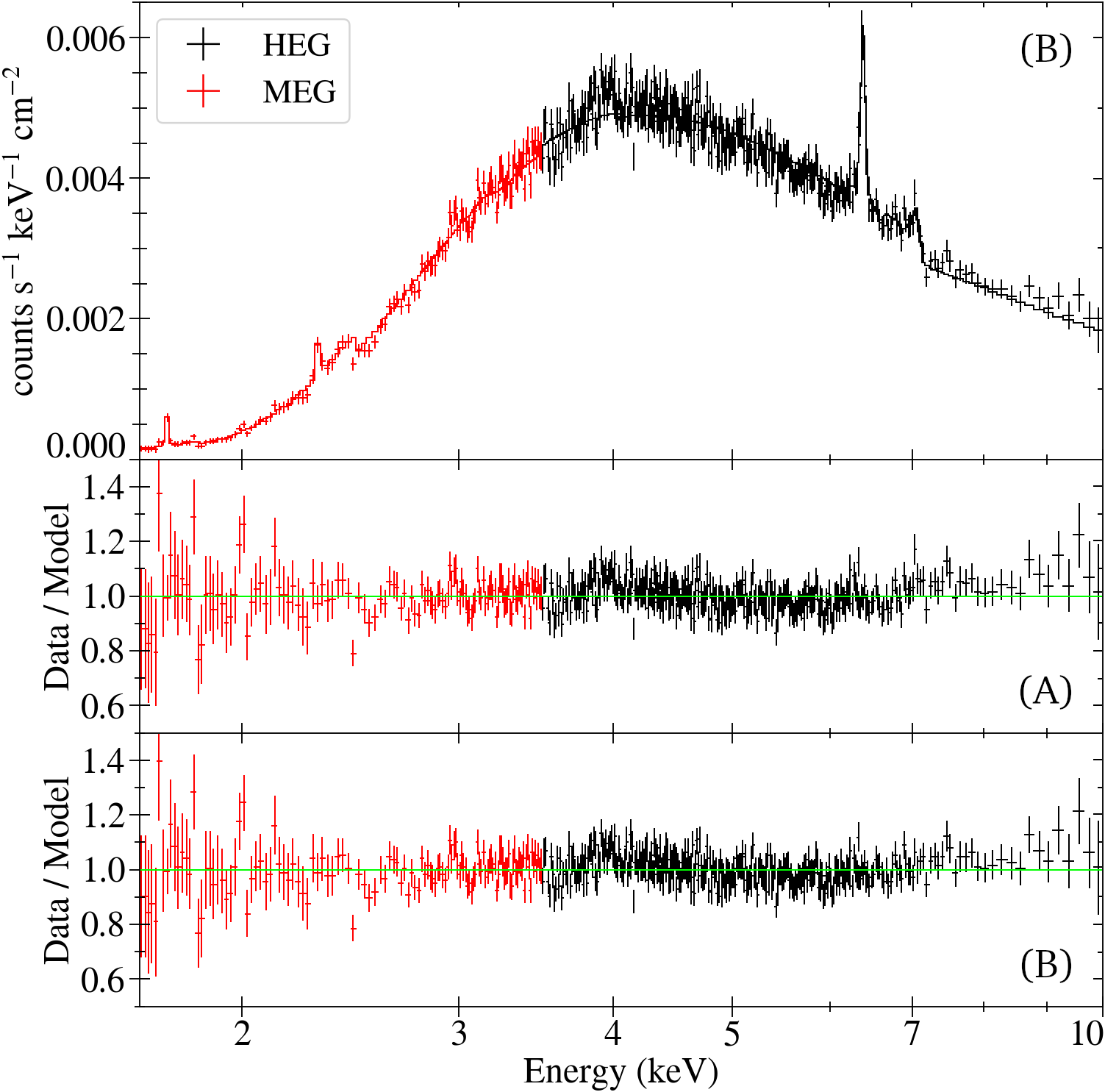}}
\caption{The best fit spectra, and the ratio of the data to the folded model, for the grouped total spectrum. The layout of the spectra is identical to that of Fig. \ref{fig:specmod01}. 
 \label{fig:specmodall}}
\end{figure}

\end{document}